\documentclass[preprint,12pt]{elsarticle}

 \usepackage{graphics}
\usepackage{epsfig}

\usepackage{amssymb}

\journal{Astroparticle Physics, in press, DOI: 10.1016/j.astropartphys.2016.04.007}

\begin{document}

\begin{frontmatter}



\title{Detection of thermal neutrons with the PRISMA-YBJ array in Extensive Air Showers selected by the ARGO-YBJ experiment}


\author{
 B.~Bartoli$^{1,2}$,
 P.~Bernardini$^{3,4}$,
 X.J.~Bi$^{5}$,
 Z.~Cao$^{5}$$^{\dagger}$,
 S.~Catalanotti$^{1,2}$,
 S.Z.~Chen$^{5}$,
 T.L.~Chen$^{6}$,
 S.W.~Cui$^{7}$$^{\dagger}$,
 B.Z.~Dai$^{8}$,
 A.~D'Amone$^{3,4}$,
 Danzengluobu$^{6}$,
 I.~De Mitri$^{3,4}$,
 B.~D'Ettorre Piazzoli$^{1,2}$,
 T.~Di Girolamo$^{1,2}$,
 G.~Di Sciascio$^{9}$,
 C.F.~Feng$^{10}$,
 Zhaoyang Feng$^{5}$,
 Zhenyong Feng$^{11}$,
 Q.B.~Gou$^{5}$,
 Y.Q.~Guo$^{5}$,
 H.H.~He$^{5}$$^{\dagger}$,
 Haibing Hu$^{6}$,
 Hongbo Hu$^{5}$,
 M.~Iacovacci$^{1,2}$,
 R.~Iuppa$^{9,12}$,
 H.Y.~Jia$^{11}$,
 Labaciren$^{6}$,
 H.J.~Li$^{6}$,
 C.~Liu$^{5}$,
 J.~Liu$^{8}$,
 M.Y.~Liu$^{6}$,
 H.~Lu$^{5}$,
 L.L.~Ma$^{5}$,
 X.H.~Ma$^{5}$\footnote[1]{Corresponding author. Tel.: +86 01088233167. E-mail address: maxh@ihep.ac.cn (Xinhua Ma)}\footnote[2]{also a member of the PRISMA Collaboration},
 G.~Mancarella$^{3,4}$,
 S.M.~Mari$^{13,14}$,
 G.~Marsella$^{3,4}$,
 S.~Mastroianni$^{2}$,
 P.~Montini$^{9}$,
 C.C.~Ning$^{6}$,
 L.~Perrone$^{3,4}$,
 P.~Pistilli$^{13,14}$,
 P.~Salvini$^{15}$,
 R.~Santonico$^{9,12}$,
 P.R.~Shen$^{5}$,
 X.D.~Sheng$^{5}$,
 F.~Shi$^{5}$,
 A.~Surdo$^{4}$,
 Y.H.~Tan$^{5}$,
 P.~Vallania$^{16,17}$,
 S.~Vernetto$^{16,17}$,
 C.~Vigorito$^{17,18}$,
 H.~Wang$^{5}$,
 C.Y.~Wu$^{5}$,
 H.R.~Wu$^{5}$,
 L.~Xue$^{10}$,
 Q.Y.~Yang$^{8}$,
 X.C.~Yang$^{8}$,
 Z.G.~Yao$^{5}$,
 A.F.~Yuan$^{6}$,
 M.~Zha$^{5}$,
 H.M.~Zhang$^{5}$,
 L.~Zhang$^{8}$,
 X.Y.~Zhang$^{10}$,
 Y.~Zhang$^{5}$,
 J.~Zhao$^{5}$$^{\dagger}$,
 Zhaxiciren$^{6}$,
 Zhaxisangzhu$^{6}$,
 X.X.~Zhou$^{11}$,
 F.R.~Zhu$^{11}$, and
 Q.Q.~Zhu$^{5}$
 \\ (The ARGO-YBJ Collaboration)
 \\
 Yu.V. Stenkin$^{19,20}$,
 V.V. Alekseenko$^{19}$,
 V. Aynutdinov$^{19}$,
 Z.Y. Cai$^{7}$,
 X.W. Guo$^{7}$,
 Y. Liu$^{10}$,
 V. Rulev$^{19}$,
 O.B. Shchegolev$^{19}$,
 V. Stepanov$^{19}$,
 V. Volchenko$^{19}$, and
 H. Zhang$^{7}$
 \\ (The PRISMA Collaboration)
 }


\address{

 $^{1}$ Dipartimento di Fisica dell'Universit\`a di Napoli
                  ``Federico II'', Complesso Universitario di Monte
                  Sant'Angelo, via Cinthia, 80126 Napoli, Italy\\
 $^{2}$ Istituto Nazionale di Fisica Nucleare, Sezione di
                  Napoli, Complesso Universitario di Monte
                  Sant'Angelo, via Cinthia, 80126 Napoli, Italy\\
 $^{3}$ Dipartimento Matematica e Fisica "Ennio De Giorgi",
                  Universit\`a del Salento,
                  via per Arnesano, 73100 Lecce, Italy\\
 $^{4}$ Istituto Nazionale di Fisica Nucleare, Sezione di
                  Lecce, via per Arnesano, 73100 Lecce, Italy\\
 $^{5}$ Key Laboratory of Particle Astrophysics, Institute
                  of High Energy Physics, Chinese Academy of Sciences,
                  P.O. Box 918, 100049 Beijing, P.R. China\\
 $^{6}$ Tibet University, 850000 Lhasa, Xizang, P.R. China\\
 $^{7}$ Hebei Normal University, Shijiazhuang 050016,
                   Hebei, P.R. China\\
 $^{8}$ Yunnan University, 2 North Cuihu Rd., 650091 Kunming,
                   Yunnan, P.R. China\\
 $^{9}$ Istituto Nazionale di Fisica Nucleare, Sezione di
                  Roma Tor Vergata, via della Ricerca Scientifica 1,
                  00133 Roma, Italy\\
 $^{10}$ Shandong University, 250100 Jinan, Shandong, P.R. China\\
 $^{11}$ Southwest Jiaotong University, 610031 Chengdu,
                   Sichuan, P.R. China\\
 $^{12}$ Dipartimento di Fisica dell'Universit\`a di Roma
                  ``Tor Vergata'', via della Ricerca Scientifica 1,
                  00133 Roma, Italy\\
 $^{13}$ Dipartimento di Fisica dell'Universit\`a ``Roma Tre'',
                   via della Vasca Navale 84, 00146 Roma, Italy\\
 $^{14}$ Istituto Nazionale di Fisica Nucleare, Sezione di
                  Roma Tre, via della Vasca Navale 84, 00146 Roma, Italy\\
 $^{15}$ Istituto Nazionale di Fisica Nucleare, Sezione di Pavia,
                   via Bassi 6, 27100 Pavia, Italy\\
 $^{16}$ Osservatorio Astrofisico di Torino dell'Istituto Nazionale
                   di Astrofisica, via P. Giuria 1, 10125 Torino, Italy\\
 $^{17}$ Istituto Nazionale di Fisica Nucleare,
                   Sezione di Torino, via P. Giuria 1, 10125 Torino, Italy\\
 $^{18}$ Dipartimento di Fisica dell'Universit\`a di
                   Torino, via P. Giuria 1, 10125 Torino, Italy\\
 $^{19}$ Institute for nuclear Research, Russian Academy of Sciences, Moscow, Russia\\
 $^{20}$ National Research Nuclear University MEPhI (Moscow Engineering Physics Institute), Moscow, Russia\\
}

\begin{abstract}
 We report on a measurement of thermal neutrons, generated by the hadronic component of extensive air showers (EAS), by means of a small array of EN-detectors developed for the PRISMA project (PRImary Spectrum Measurement Array), novel devices based on a compound alloy of ZnS(Ag) and $^{6}$LiF.  This array has been operated within the ARGO-YBJ experiment at the high altitude Cosmic Ray Observatory in Yangbajing (Tibet, 4300 m a.s.l.). Due to the tight correlation between the air shower hadrons and thermal neutrons, this technique can be envisaged as a simple way to estimate the number of high energy hadrons in EAS. Coincident events generated by primary cosmic rays of energies greater than 100 TeV have been selected and analyzed. The EN-detectors have been used to record simultaneously thermal neutrons and the air shower electromagnetic component. The density distributions of both components and the total number of thermal neutrons have been measured. The correlation of these data with the measurements carried out by ARGO-YBJ confirms the excellent performance of the EN-detector.
\end{abstract}


\end{frontmatter}


\section{Introduction}

The cosmic ray energy spectrum spans over many decades from about 10$^6$ eV to beyond 10$^{20}$ eV. It consists of different regions with power law behavior and changes in the power law index.  In the high energy range above 100 TeV  two features are known since a long time, that is a steepening of the spectrum, named the knee, at about 3-5 $\times$ 10$^{15}$ eV and a hardening, named the ankle, at about 3-5 $\times$ 10$^{18}$ eV . Other peculiar features have been observed in this energy interval by the KASCADE-Grande experiment \cite{K-G}. The hybrid experiment ARGO/WFCTA  has recently established a bending of the light component (protons and Helium nuclei) at about 700 TeV \cite{A-W}. Although the global features of the all-particle spectrum are reasonably recognized, the spectral shape of each primary component remains an open question related to the interpretation of the experimental data. On the other hand, the determination of the energy spectrum and chemical composition of cosmic rays bears important information on their origin, acceleration and propagation mechanisms. The direct observation of this radiation is accomplished with high efficiency by means of balloon-borne detectors or equipments installed on satellites. However, due to the limited pay-load, these measurements are constrained by the small exposure. Because of the steeply falling cosmic ray spectrum, the detected rate drops quickly leaving only a few events per year detectable at energies approaching the PeV range. The investigation of the low fluxes of high energy cosmic rays is addressed by means of ground-based experiments, where many detectors are deployed over large areas. Indeed, when arriving at Earth, high energy cosmic rays interact with the air nuclei originating extensive air showers (EAS). They consist of a core of high energy hadrons that continuously feed the electromagnetic part of the shower, mainly with photons from neutral pion, kaon and eta particle decays. Nucleons and other high energy hadrons contribute to the hadronic cascade, while the decays of low energy unstable secondaries, as charged pions and  K-mesons, generate muons and neutrinos. Other neutrinos come from the decay of low energy muons. Thus, the EAS develop in atmosphere as an avalanche process in three different components which are, in decreasing order of intensity, the electromagnetic, the muon and the hadronic component. The intensities of these components are strictly correlated, with a correlation depending on the energy and nature of the primary particle and on the stage of the cascade process.  Cherenkov radiation, atmospheric scintillation light and radio emission accompany the development of the atmospheric particle cascade. All these particles are sampled at ground level, while the so-called penetrating component, that is high energy muons and neutrinos, are usually recorded in underground detectors. Inferring the energy and nature of the primary particle from the measurement of a single EAS component is a very hard task, often requiring some a priori model concerning energy spectrum and chemical composition \cite{Tibet2}\cite{ICE-TOP2}\cite{MAKET2}. In modern experiments a multiparametric approach, based on the simultaneous detection of some of the EAS observables and their correlation, is carried out to infer the features of the cosmic ray spectrum. In addition to the electromagnetic component, muons (as in the KASCADE \cite{KASCADE}, CASA-MIA \cite{CASA-MIA}, EAS-TOP \cite{EAS-TOP} and ICE-TOP \cite{ICE-TOP} experiments) or Cherenkov light (as in DICE \cite{CASA-DICE} and ARGO/WFCTA \cite{A-W}\cite{ARGO3}) are the most common EAS observables used for this purpose. Some specific experimental arrangements, as in the Tibet AS experiment \cite{ASg}, which uses burst detectors to sample high energy photons, can be also implemented. On the contrary, an extensive use of the hadron detection to get information on the cosmic ray spectrum was limited up to now by the absence of simple and cheap hadron detectors, being very expensive and quite complicated the use of conventional hadron calorimeters (HCAL) over large areas \cite{EAS-TOP2}\cite{KASCADE3}\cite{KASCADE4}.  Emulsion chambers (EC) have been also used to detect high energy hadrons, mainly in stand-alone experiments at high altitude \cite{MEC1}\cite{MEC2}\cite{MEC3}\cite{MEC4} . Their use in air shower experiments is not straightforward, requiring complex procedures to associate the EC events with the air showers that caused them \cite{Matano}. On the other hand, high energy hadrons, which constitute the EAS skeleton, may carry important information for multi-parameter correlation studies, since some hadronic observables, primarily  the  hadron number/electron number correlation, depend on the nature of the particle inducing the shower \cite{PRISMA1}\cite{PRISMA6}. In particular, the hadron content is an excellent mark to select photon-induced showers. Thus, the detection of high energy hadrons, addressed to improve the discrimination power in these analysis, is highly advisable.

A way to deal with this problem avoiding the use of huge and expensive HCALs was brought out in \cite{PRISMA1}\cite{PRISMA2}. In these papers the detection of thermal neutrons generated by EAS hadrons is proposed. It is well known that hadrons interacting with ambient matter (air, building, ground, etc.) produce evaporation neutrons due to nuclei disintegration. The neutrons have no charge and lose energy only by scattering. If the medium is a good moderator, i.e., the absorption cross section is much less than the scattering cross section, the neutrons lose energy via scattering down to the thermal ones (moderation process) and then live in the matter until capture. Evaporation neutrons need about 0.5 ms to thermalize in rock (concrete). Neutrons are generated abundantly, up to 2 orders of magnitude more than parent hadrons. The mean number of evaporation neutrons $<$n$>$ produced by hadrons in a 120 cm layer of surrounding soil (about 3 hadron interaction lengths) and/or construction materials can be estimated using the empirical relationship
\begin{equation}
<n> {\approx} ~36 {\times} E^{0.56}_{h}
\label{nE}
\end{equation}
where $E_h$ is the hadron energy in GeV. The relation was obtained using the experimental data on neutron production for $E_h$ $>$ 5 GeV \cite{neutron7} and then recalculated to soil taking into account dependence on target atomic number \cite{neutron72}. Since the energy distribution of hadrons in EAS exhibits a very slow dependence on the primary energy, the total number of evaporation neutrons is expected to be proportional to the total number of high energy hadrons reaching the observation level \cite{PRISMA3}. A large fraction of the evaporation neutrons thermalize, so that recording thermal neutrons can be exploited to reconstruct the hadron content in the shower. This approach looks very promising for measurements carried out at high altitude. Indeed, since the hadron content in EAS increases with the altitude, an abundant production of thermal neutrons  can be predicted for  experiments at 4 (or more) km a.s.l. , about a factor 10 higher than that at sea level \cite{PRISMA3}. These considerations suggested the development of a simple and cheap thermal neutron detector, to be deployed over a large area,  as  'hadron counter' in EAS experiments at mountain level. This idea led to the development of the EN-detector, made of a mixture of the well-known inorganic scintillator ZnS(Ag) with $^{6}$LiF, capable of recording both thermal neutrons and charged particles \cite{PRISMA4}\cite{PRISMA5}.  The study of neutrons in EAS was started in the 1930s \cite{neutron1} and experiments in the late Forties succeeded in demonstrating the production of neutrons in the nuclear interaction of cosmic rays \cite{neutron2}\cite{neutron3}\cite{neutron4}. With the appearance of neutron monitors, invented by John A. Simpson in 1948, developed primarily to measure the intensity of solar cosmic rays, some  air shower experiments operated them to study the hadrons in EAS \cite{neutron7}\cite{neutron5}\cite{neutron6}\cite{neutron8}\cite{neutron9}. However neutron monitors, which usually employ $^{10}$BF$_3$ or $^3$He gas proportional counters, are not suitable for an extensive use over large areas. On the contrary, the EN-detectors, relatively simple, compact and cheap, can be  easily deployed in an air shower array to record simultaneously thermal neutrons and the charged particles in the shower front \cite{PRISMA6} . A prototype array of 32 EN-detectors (PRISMA-32) is now running in Moscow \cite{PRISMA7}\cite{PRISMA9}. In order to check the performance of this detector at a high altitude site, a small array composed of four EN-detectors (PRISMA-YBJ) has been installed inside the  hall hosting the ARGO-YBJ experiment at the Yangbajing Cosmic Ray Observatory (Tibet, China, 4300 m a.s.l. , 606 g/cm$^2$).  The two arrays operated together for about two weeks, and coincident events have been analyzed to gather information on the PRISMA-YBJ performance. Sect. 2 of this paper introduces the EN-detector, Sect. 3 gives an overview of the experimental set-up. Sect. 4 describes the coincident event selection. Data analysis and results are presented and discussed in Sect. 5. Summary and conclusions are given in Sect. 6.

\section{The EN-detector}

\begin{figure}[t]
 \centering
 \includegraphics[width=0.35\textwidth]{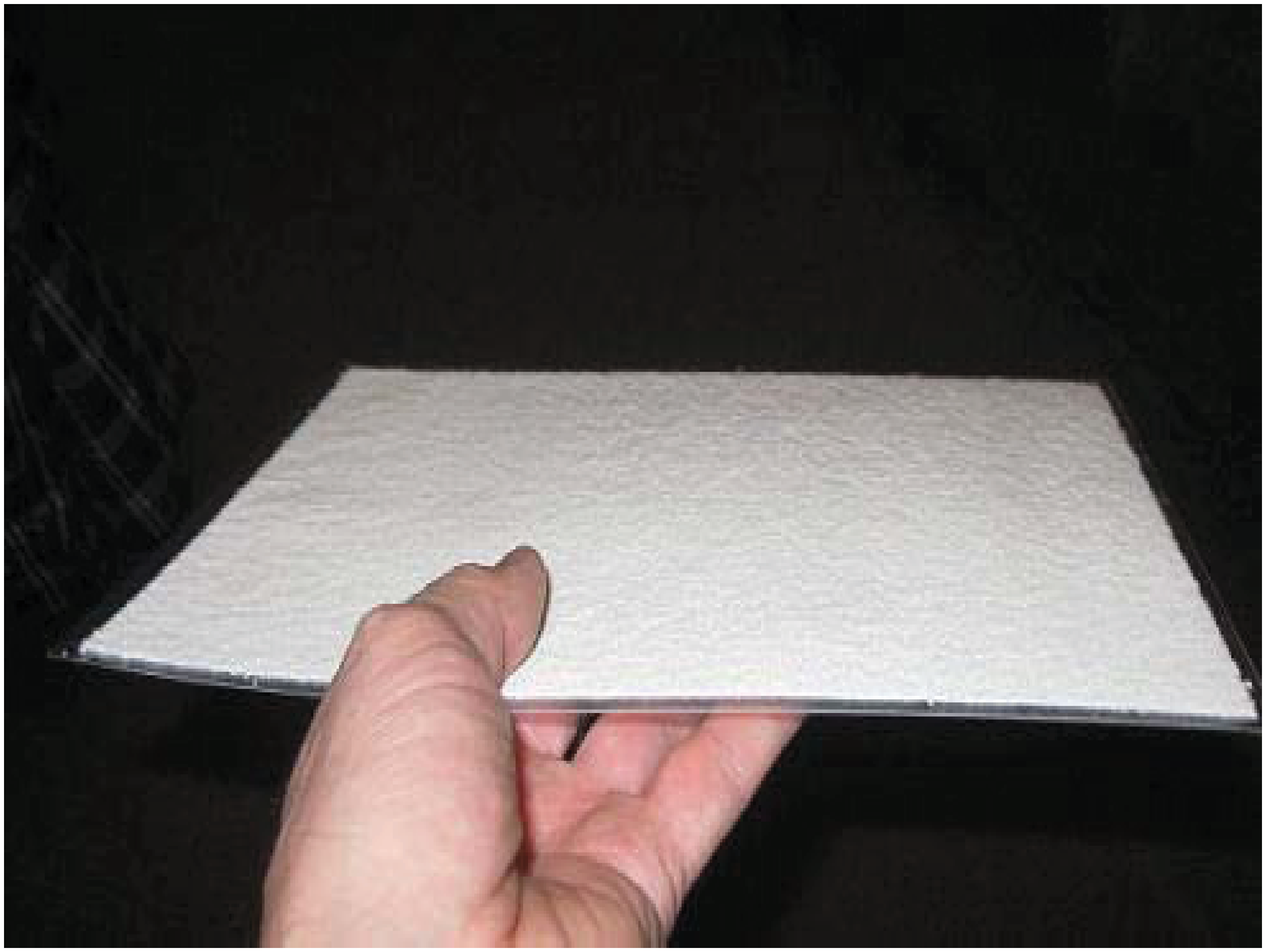}
 \includegraphics[width=0.5\textwidth]{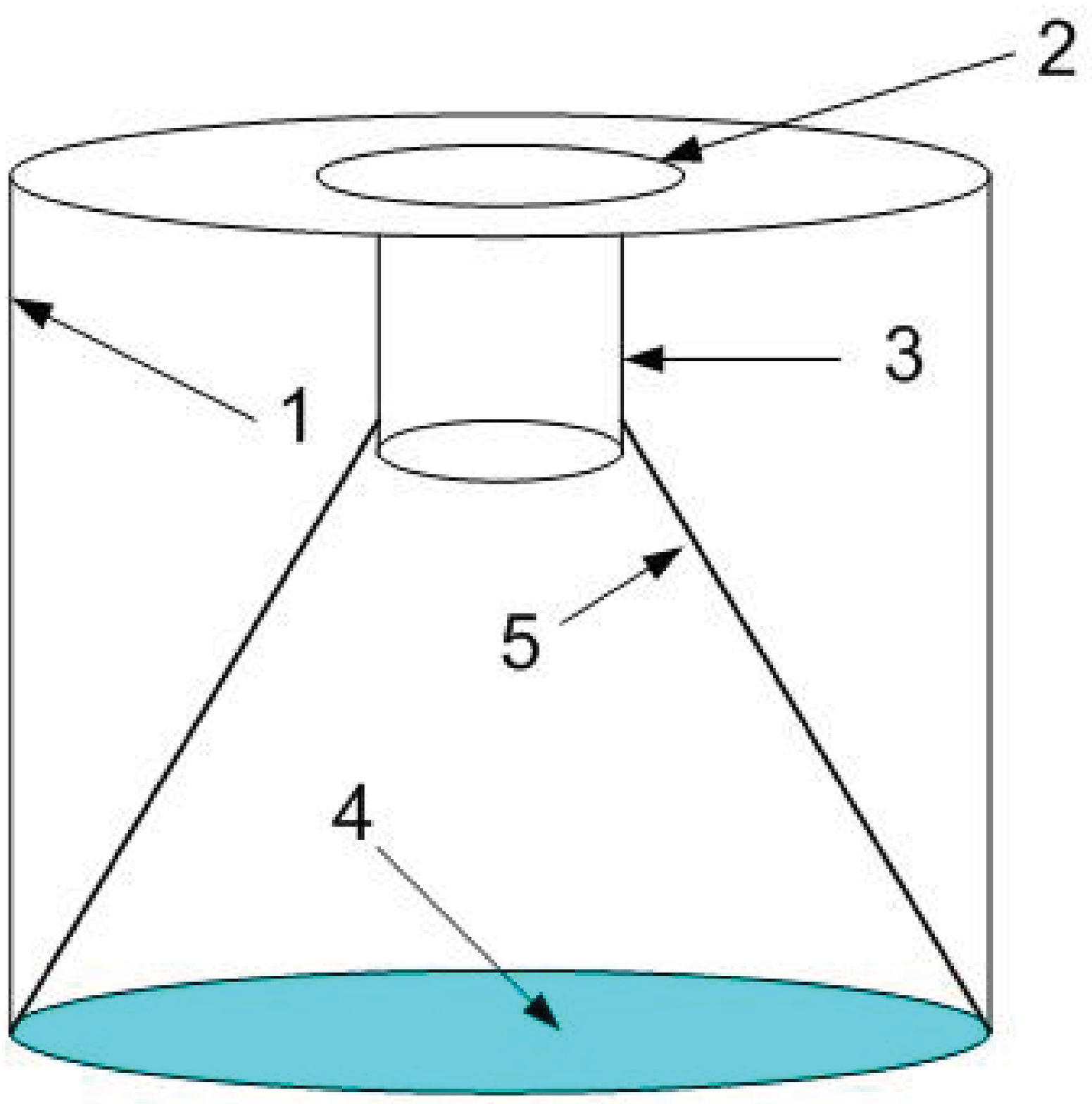}
 \caption{Left: Photo of the ZnS(Ag) scintillator. Right: Scheme of the EN-detector. 1) polyethylene tank (diameter=72 cm, height=57 cm). 2) 30 cm diameter lid. 3) 6" PMT. 4) scintillator with area 0.36 $m^2$. 5) reflecting cone.}
 \label{EN-detector}
 \end{figure}

\begin{figure}[htb!]
 \centering
 \includegraphics[width=0.7\textwidth]{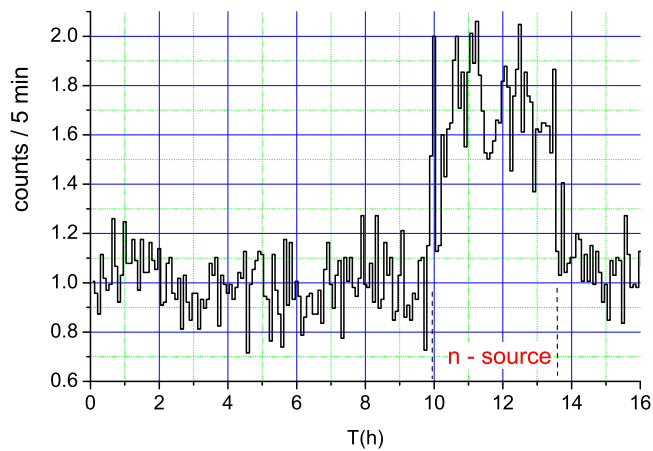}
 \caption{Neutron counting rate with and without the neutron source $^{252}Cf$. The horizontal axis gives the time in hours, the vertical axis the neutron counts in bins of 5 minutes. The background is from radiation in the surrounding environment.}
 \label{n_source}
 \end{figure}

The EN-detector is based on a special phosphor, which is a granulated alloy of inorganic ZnS(Ag) scintillator added with LiF enriched with the $^{6}$Li isotope up to 90\% \cite{PRISMA8} (Fig. \ref{EN-detector}, left). One $^{6}$Li captures one thermal neutron via the reaction $^{6}Li + n \rightarrow ^{3}H + \alpha + 4 .78 MeV$ with cross section of 945 barn. The phosphor is deposited in the form of a thin one-grain layer on a white plastic film, which is then laminated on both sides with a thin transparent film. The scintillating compound grains used are of 0.3 - 0.8 mm in size. The effective thickness of the scintillator layer is 30 $mg/cm^2$. Light yield of the scintillator is $\sim 160,000$ photons per neutron capture. The structure of a typical EN-detector is shown in Fig. \ref{EN-detector}, right. The scintillator of 0.36 $m^2$ area is mounted inside a black cylindrical polyethylene (PE) 300-l tank which is used as the detector housing. The scintillator is supported inside the tank to a distance of 36 cm from the photomultiplier (PMT) photocathode. A 6"-PMT (FEU-200) is mounted on the tank lid. A light reflecting cone made of foiled PE foam of 5-mm thickness is used to improve the light collection. As a result, $\sim$ 100 photoelectrons per neutron capture are collected. The efficiency for thermal neutron detection in our scintillator was found experimentally by neutron absorption in the scintillator layer to be about 20\%. To determine it, we measured the counting rate of our scintillator layer, then we put another similar layer under the first one (with a black paper between them) as an absorber and measured again. Then we compared the results and calculated the scintillator efficiency. Similar efficiency was also obtained by simple Monte-Carlo simulation using GEANT4 code. As an example, we show in Fig. \ref{n_source} the response of the detector illuminated with a low activity source of thermal neutrons ($\sim$1 Bq of $^{252}Cf$).

\begin{figure}[t]
 \centering
 \includegraphics[width=0.8\textwidth]{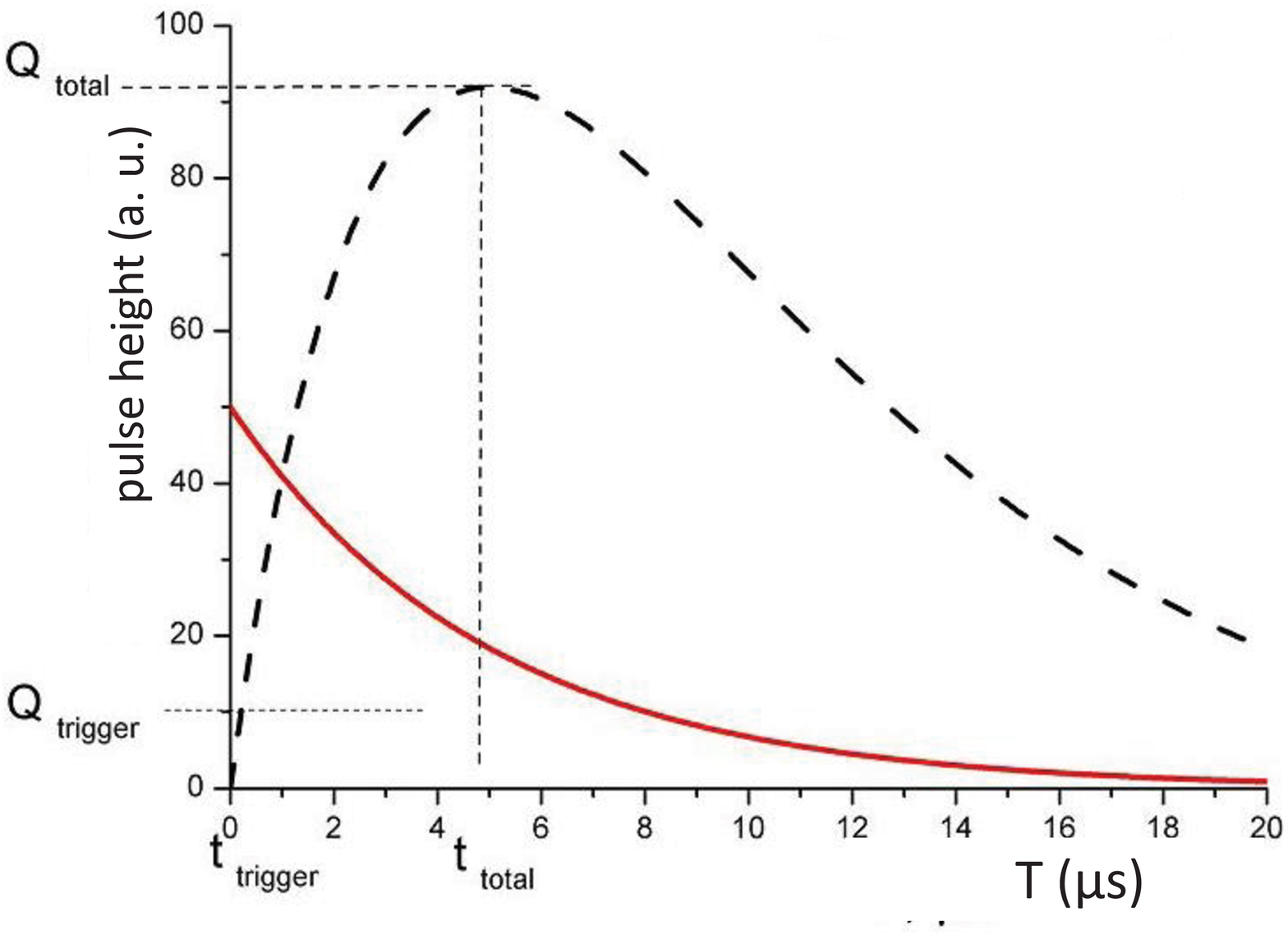}
 \caption{Different pulse shapes of neutrons (dashed line) and electrons (from a small size EAS). $Q_{trigger}$ and $t_{trigger}$ are charge and time when the pulse passes the threshold, respectively. $Q_{total}$ and $t_{total}$ are charge and time when the pulse height reaches the peak. (For interpretation of the references to color in this figure legend, the reader is referred to the web version of this article.)}
 \label{alpha-e}
 \end{figure}

The EN-detector is sensitive to charged particles as well as to thermal neutrons. However the light output is different for different types of particles. This characteristic makes possible to select neutron signals from those generated by charged particles (or gammas) exploiting their different amplitude and pulse shape. Due to the thin layer of the scintillator, charged particles deposit on average only 60 keV against 4.8 MeV deposited during the neutron capture. A very high alpha/e ratio, that is the ratio of the light produced by alpha particles to the light produced by electrons of the same energy, is the main detector feature. This feature allows, in principle, to distinguish with high efficiency different types of particles. However, in environments of intense natural radioactivity in addition to the cosmic ray background, the tail of the distribution of the energy deposited by charged particles could mimic a neutron signal. The different pulse shape of the neutron signal with respect to the signal produced by charged particles can be fruitfully exploited to remove this background. Indeed, slowly moving heavy particles (such as alphas or tritons) excite slow components in addition to the emission of fast signals. The charge collection time of a signal due to a neutron capture recorded at the PMT 12th dynode of the EN-detector  is 10-20 $\mu$s , while the characteristic time of the fast emission induced by charged particles is about 40 ns.  We compare in Fig. \ref{alpha-e} the pulse shape of the neutron signal with the signal induced by electrons. The remarkable difference in shape allows an efficient use of  pulse-shape discrimination to select and record neutron signals in measurements of a neutron flux. Note that all signals are digitized with a FADC whose resolution is equal to 1 V / 1024 ch = 1 mV/ch . A preliminary absolute calibration on the EAS electromagnetic component of the EN-detectors operated at PRISMA-YBJ experiment has been performed in Moscow using the NEVOD experimental facility in MEPhI by means of a small EAS array of plastic scintillators selecting atmospheric showers \cite{PRISMA11}. The EN-detectors have been installed between plastic scintillators ($\sim$ 1 m spacing) and their response to shower particles allowed us to determine with an accuracy of $\pm$ 20\%  the relation between the charge delivered by the PMTs (in FADC channels) and the number of particles crossing the EN-detector \cite{PRISMA10}. Obtained results gave us the calibration value of 2.44 $\pm$ 20\% minimum ionizing particles (m.i.p.s)/channel. This coefficient was used to recover the particle density in the PRISMA-YBJ experiment (see Sect. 5). Our threshold for data acquisition is 1 mV corresponding to about 2.5 particles per detector. Only in the case of synchronous passage of at least 3 particles through the detector a signal suitable for recording is generated. This feature allows one to use the EN-detector in scaler mode for low neutron flux measurements.

\begin{figure}[t]
 \centering
 \includegraphics[width=0.8\textwidth]{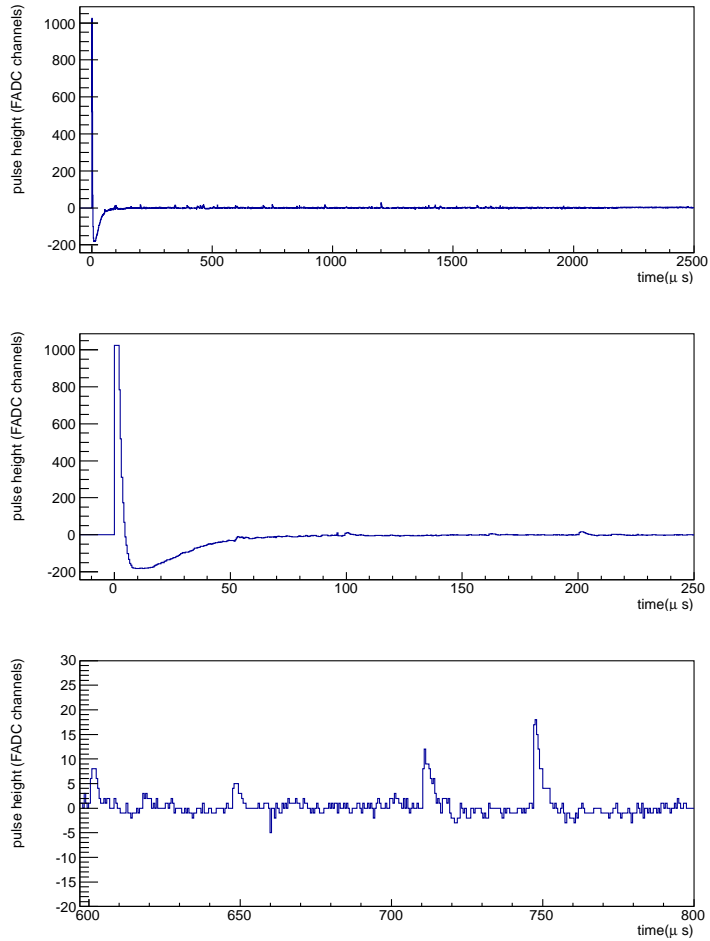}
 \caption{The shape of the signals from the neutron detectors in coincidence with an EAS event recorded by ARGO-YBJ. Upper plot: the pulse from 0 to 2.5 ms. The large peak in the first bin is generated by the EAS electrons. Middle plot: the pulse from 0 to 0.25 ms. Lower plot: the pulse from 0.6 to 0.8 ms (note the different scale on the vertical axis). The small peaks following the first peak are generated by thermal neutrons.}
 \label{easpulse}
 \end{figure}

The peculiar characteristics of the EN-detector output, that are weak and fast signals from charged particles compared to high amplitude, slow and delayed signals from thermal neutron capture, make it well suitable for its use in the framework of EAS experiments. In high energy EAS the time thickness of the shower front is about tens of ns , depending on the distance from the core. The individual signals generated by these particles (mainly electrons and positrons) add up to give a signal proportional to their number which can be used also for triggering and timing purposes. Delayed signals from thermal neutron capture follow on a time scale of a few milliseconds. As an example,we show in Fig. \ref{easpulse} the pulses recorded in a high energy EAS event. The first big peak is generated by the large amount of charged particles of the shower front while the  smaller delayed signals are generated by thermal neutrons. Thus, the amplitude of the fast signal can be used to measure the charged particle density while the delayed signals measured in a time gate of 10 ms give the number of captured thermal neutrons. The selection of electrons and neutrons is automatically performed by the off-line analysis program.

\section{The Experimental Set-up}

ARGO-YBJ \cite{ARGO1} is a full coverage air shower detector constituted by a central carpet 74 $\times$ 78 $m^2$, made of a single layer of RPCs (2.85 $\times$ 1.23 $m^2$ each) with 93\% active area, enclosed by a guard ring partially instrumented (about 20\%)up to 100 $\times$ 110 $m^2$ . The apparatus has a modular structure, the basic data acquisition element being a cluster (5.7 $\times$ 7.6 $m^2$) made of 12 RPCs. The chambers are operated in streamer mode. Both digital and charge read-out are implemented by means of external electrodes. The digital read-out is realized by means of strips (6.75 $\times$ 61.80 $cm^2$) which provide signals used for timing and to build up the trigger. An inclusive trigger is implemented , exploiting a majority logic, which allows  the detection of small size showers . More details about the detector and the RPC performance can be found in Ref. \cite{ARGO4}. Because of the small pixel size, the detector is able to record events with a particle density exceeding 0.02 $particles/m^2$ keeping good linearity up to a density of about 15 $ particles/m^2$ . This output has been used to measure the  light component spectrum (protons+Helium nuclei) of the primary cosmic rays up to 300 TeV (\cite{ARGO5}) and to carry out many observations in gamma-astronomy (Ref. \cite{ARGO6} and ref. therein) . In order to extend the dynamical range up to PeV energies,each chamber is also equipped with two large size pads (BigPads, 139 $\times$ 123 $cm^2$ each) to collect the total charge developed by the particles hitting the detector\cite{ARGO7}. In this way the measurement of  densities up to $10^4$ $particles/m^2$  can be achieved with an accuracy better than 10\% \cite{ARGO8}\cite{ARGO9}. The read-out of the charge collected  by  the BigPads is enabled in each cluster by a local trigger based on the particle multiplicity. 100\% trigger efficiency above 100 TeV for all kind of primary particles is achieved. The use of the analog read-out has allowed the measurement of the light component spectrum up to 700 TeV\cite{ARGO3}, and this measurement has been extended to the PeV region \cite{A-W}. The event arrival time is measured by GPS (Global Positioning System) with precision of 100 ns. During the common runs with PRISMA-YBJ, the ADC full scale of the BigPads has been set so as to sample with full efficiency showers induced by primaries with energies from about 100 TeV to many PeV. ARGO-YBJ accurately measures the showers, to obtain several parameters including shower direction, core position, age and size.

\begin{figure}[t]
 \centering
 \includegraphics[width=0.8\textwidth]{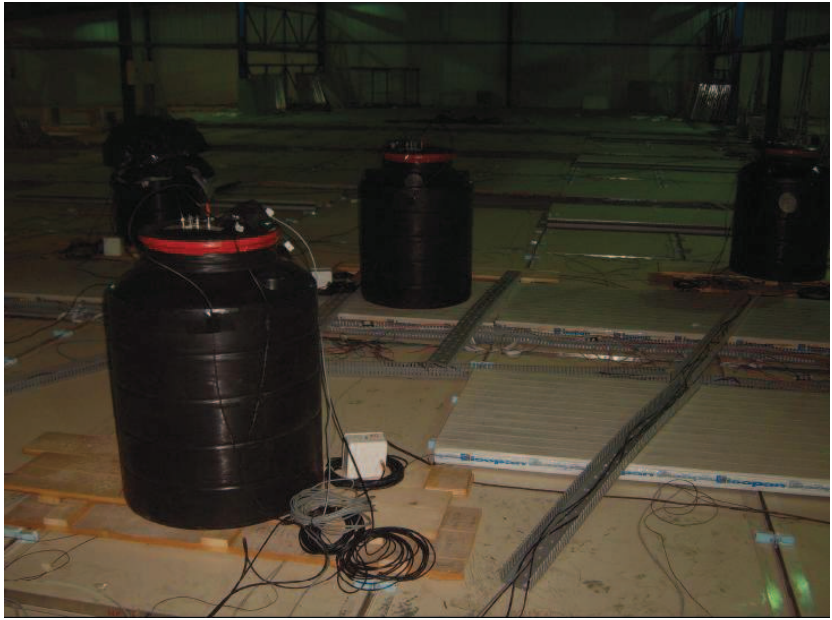}
 \includegraphics[width=1\textwidth]{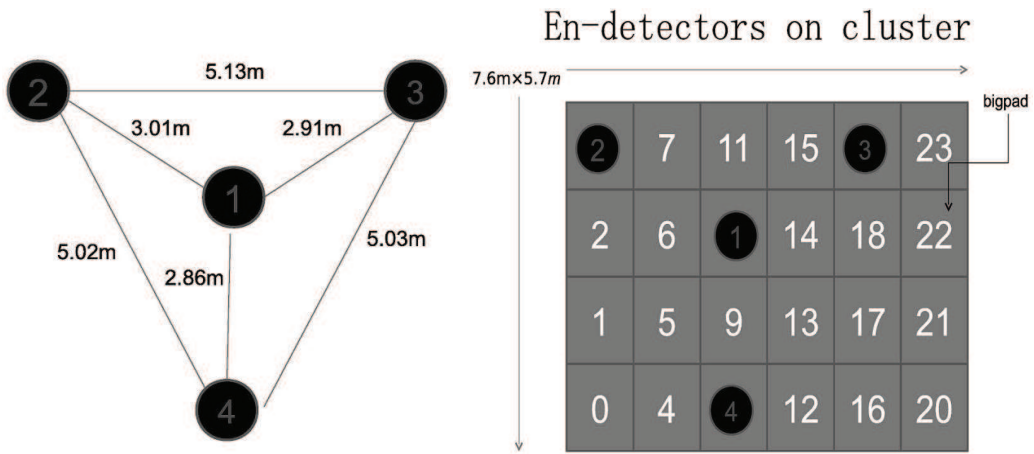}
 \caption{Upper plot: Photo of PRISMA-YBJ above the ARGO-YBJ RPC carpet. Lower left: Layout of PRISMA-YBJ. Lower right: PRISMA-YBJ on the ARGO-YBJ cluster 78.}
 \label{array}
 \end{figure}

The four EN-detectors of the PRISMA-YBJ array have been installed in the hall hosting the ARGO-YBJ experiment. Three of them are located according to a triangular arrangement on one of the ARGO-YBJ clusters (cluster No. 78), each side of the triangle being about 5m. The fourth EN-detector is installed at the center of the triangle at a distance of about 3 m from the other detectors. Each EN-detector overlaps one of the Big Pads of the cluster (Fig. \ref{array})

\begin{figure}[t]
 \centering
 \includegraphics[width=1\textwidth]{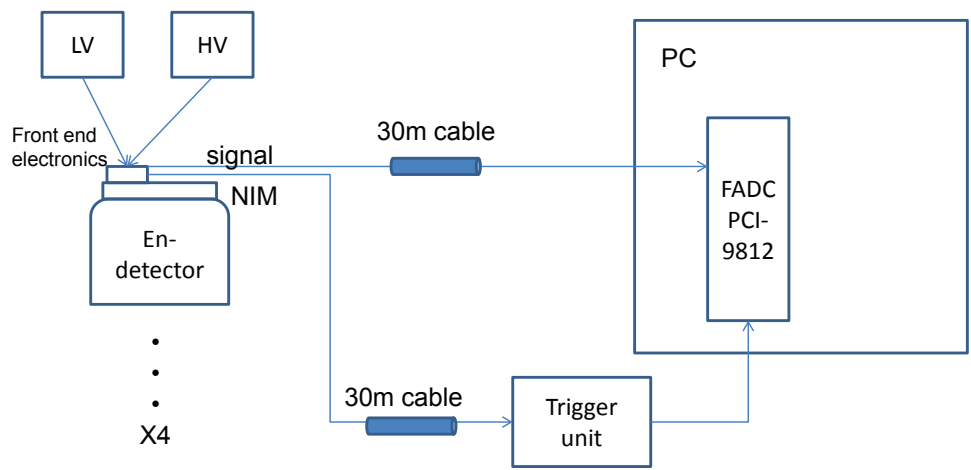}
 \caption{Scheme of PRISMA-YBJ electronics.}
 \label{scheme_elec}
 \end{figure}

A schematic view of the recording system of PRISMA-YBJ is shown in Fig. \ref{scheme_elec}. The signal from the 12th dynode of each PMT is sent to a charge sensitive preamplifier-discriminator where it is split into two pulses: one of them is shaped to a NIM pulse used to build-up the trigger, the other one is integrated with a 1 $\mu$s time constant, then amplified and sent to the input of a FADC (ADLINK 12 PCI-9812). The first pulse produced mostly by EAS electrons is used for trigger and energy deposit measurements, and delayed neutron capture pulses are counted within a time gate of 10 ms to give the number of neutrons. The first level trigger which is a coincidence of any 2 out of 4 detectors in a time gate of 1 $\mu$s,  starts all FADC's. The on-line program analyzes the input data and provides the following second level triggers:
 \par M1, if at least 2 detectors generating the first level trigger measured a signal corresponding to 10 m.i.p.s or more;
 \par M2, if the delivered total charge corresponds to more than 125 m.i.p.s;
 \par
\begin{figure}[t]
 \centering
 \includegraphics[width=1\textwidth]{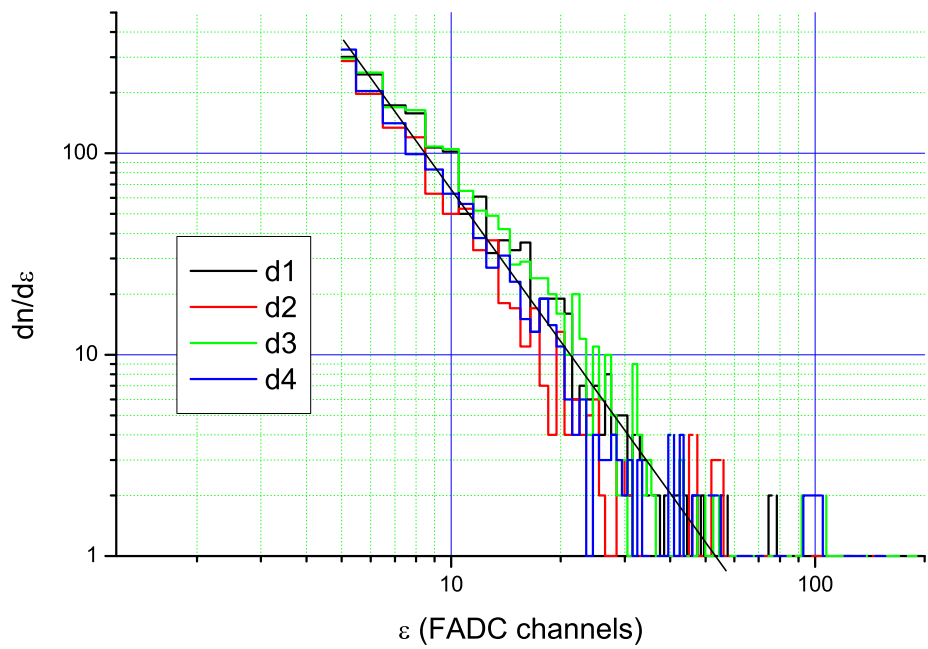}
 \caption{Daily charge spectra of the EN-detectors produced by relativistic particles in events triggered by the first level trigger. The horizontal axis gives the charge $\epsilon$ delivered by the FADCs.}
 \label{monitor}
 \end{figure}

 M3, if the total number of recorded neutrons is higher than 4.\\
If any of these conditions is fulfilled thus the pulse shape of all signals is stored (20000 samples in steps of 0.5 $\mu$s.)  along with the mark of the trigger number. In addition, every 5 minutes the on-line program generates a trigger (M0) which starts the data acquisition by the FADCs. This 'random' trigger, not related to showers, allows the measurement of chance signals which could mimic neutron signals in the 10 ms recording window. The EN-detectors were routinely monitored recording the charge spectra accumulated by each detector. An example is given in Fig. \ref{monitor} which shows the daily spectra produced by relativistic particles in events triggered by the first level trigger. We observe a satisfactory equalization of the detector gain. The event arrival time is measured by the network timing protocol (NTP) method.  NTP is a networking protocol for continuous clock synchronization of computer systems to internet time server. NTP can provide timing with a precision of $\sim$ 10 ms to Coordinated Universal Time (UTC). PRISMA-YBJ and ARGO-YBJ were triggered by EAS events individually and coincident events were searched for during off-line data analysis.

\section{Coincident EAS events between ARGO-YBJ and PRISMA-YBJ }

\begin{figure}[t]
 \centering
 \includegraphics[width=0.7\textwidth]{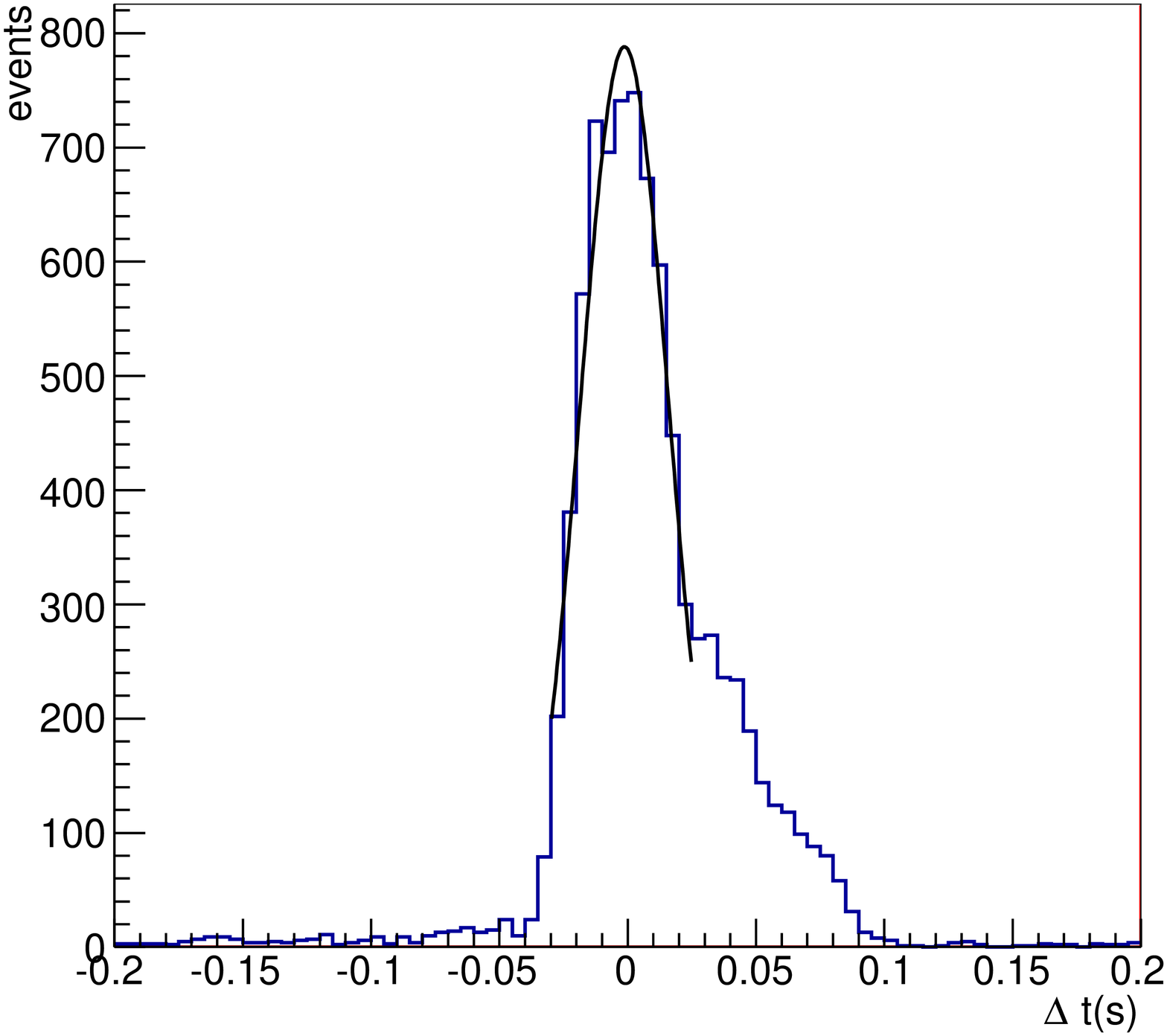}
 \includegraphics[width=0.7\textwidth]{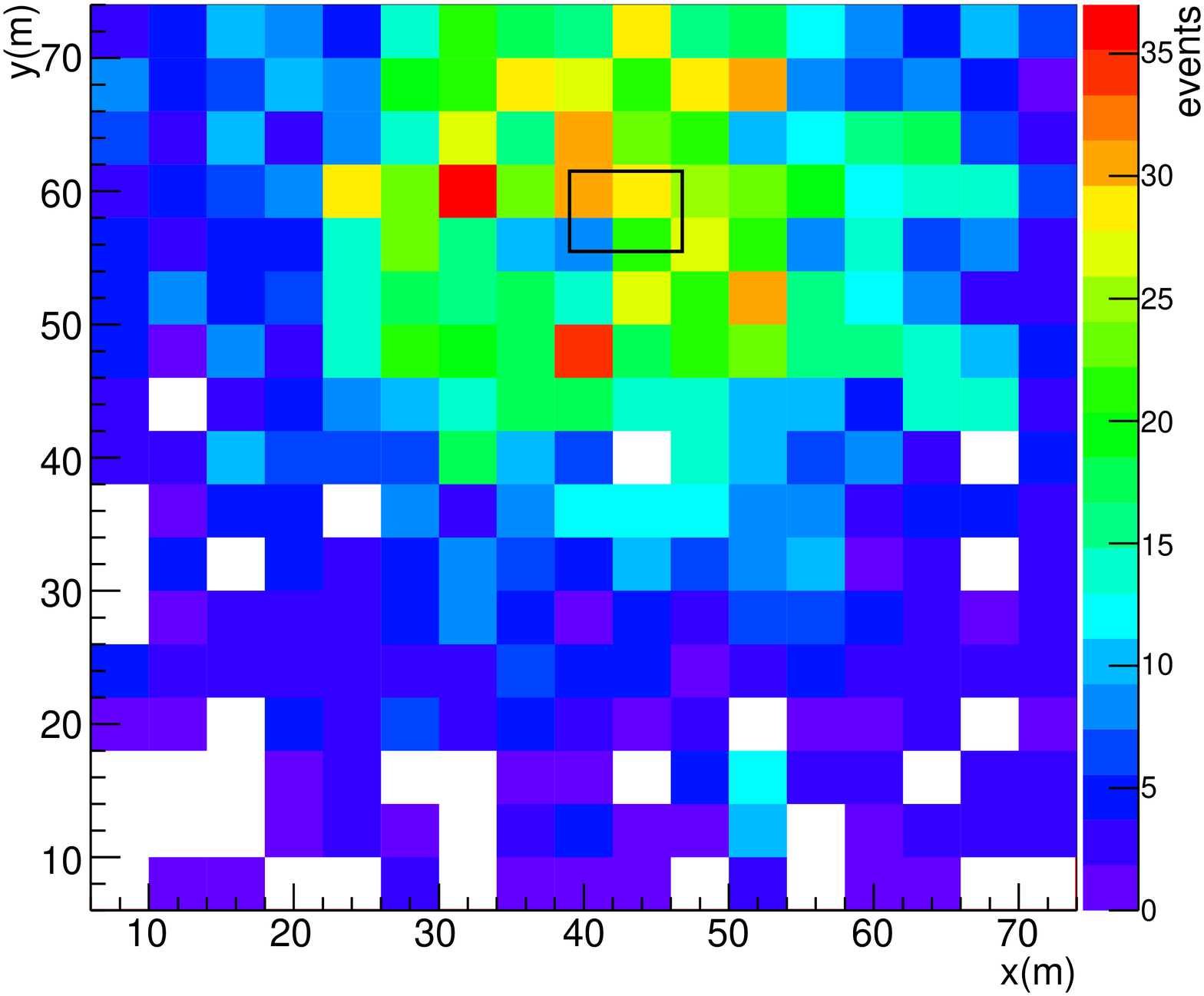}
 \caption{Upper plot: Distribution of the event time differences between PRISMA-YBJ and ARGO-YBJ fitted with a Gaussian function. The mean value is -0.0013 $\pm$ 0.0003, the s.d. is 0.01716 $\pm$ 0.00032. Lower plot: Distribution of the core positions of the matched events. The black rectangle represents cluster 78. (For interpretation of the references to color in this figure legend, the reader is referred to the web version of this article.)}
 \label{selection}
 \end{figure}

\begin{figure}[t]
 \centering
 \includegraphics[width=0.8\textwidth]{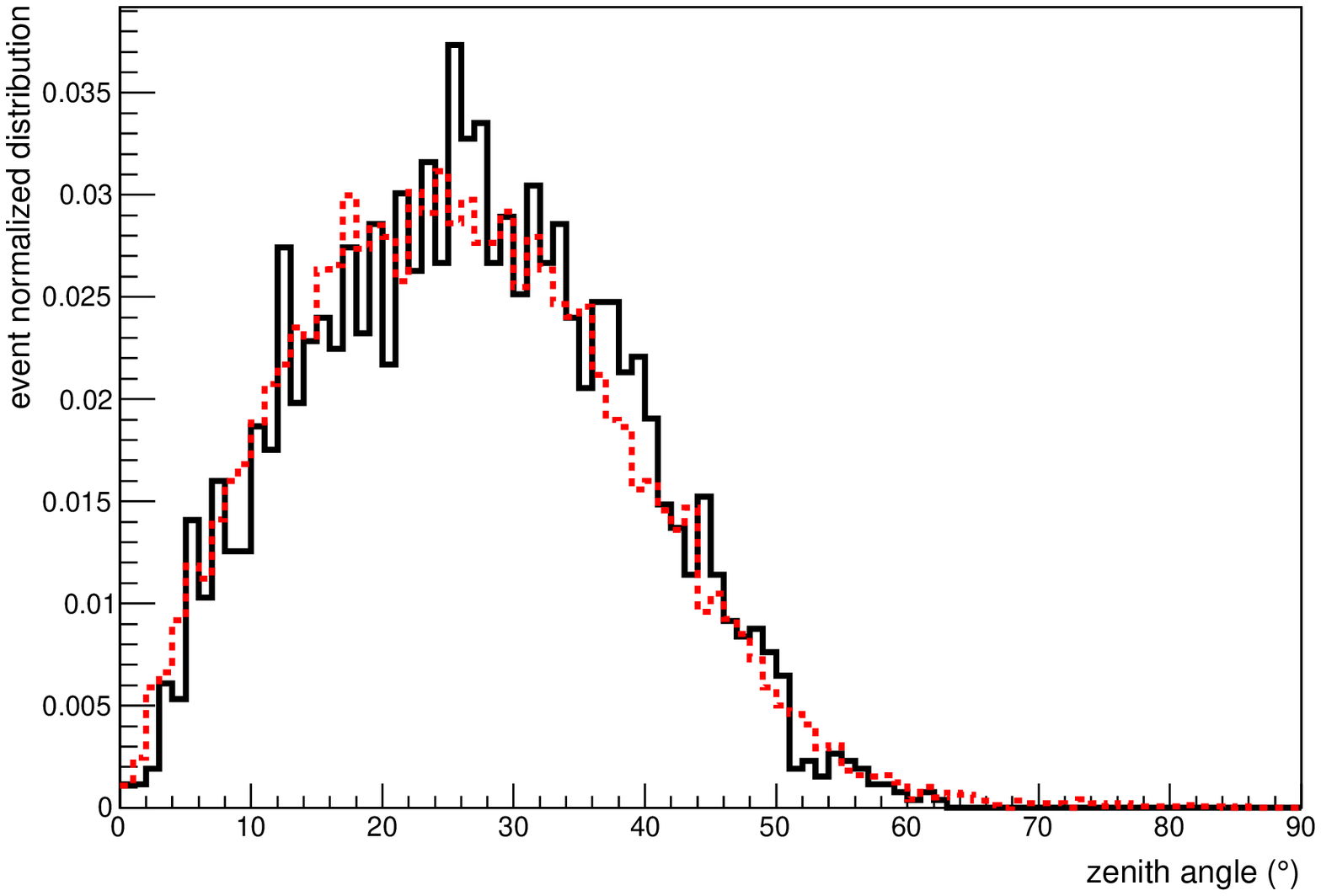}
 \includegraphics[width=0.8\textwidth]{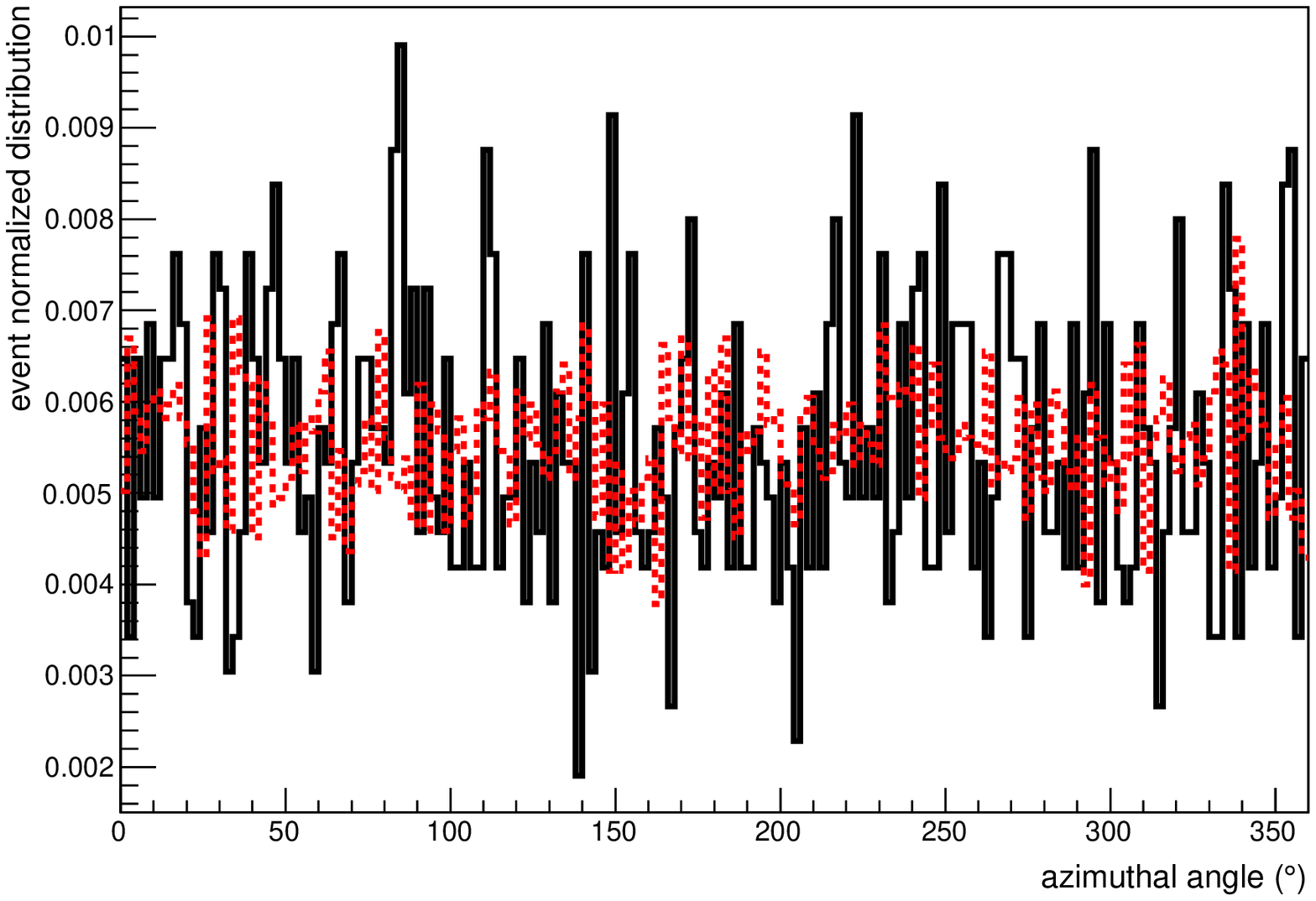}
 \caption{Upper plot: Zenithal distribution of the coincident (black solid line) and ARGO-YBJ (red dashed line) events. Lower plot: Azimuthal distribution of the coincident (black solid line) and ARGO-YBJ (red dashed line) events. Both distributions are normalized to the corresponding total number of events. (For interpretation of the references to color in this figure legend, the reader is referred to the web version of this article.)}
 \label{angular}
 \end{figure}

\begin{figure}[t]
 \centering
 \includegraphics[width=0.58\textwidth]{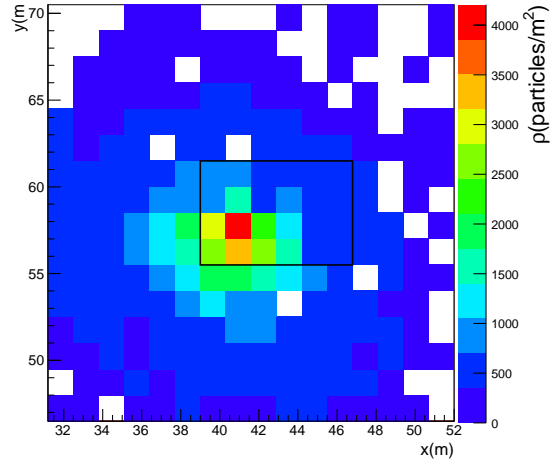}
 \includegraphics[width=0.58\textwidth]{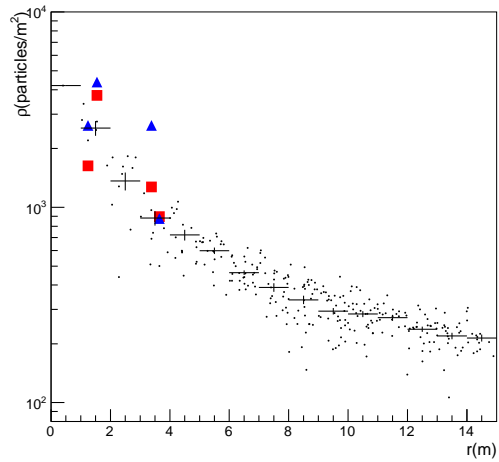}
 \caption{Upper plot: Scatter plot of charged particle density measured by ARGO-YBJ for a coincident event, where each pixel gives the charge delivered by the BigPad. The black rectangle represents cluster 78. Lower plot: Lateral distributions of the recorded particles for the same coincident event. Black dots: charged particle density measured by ARGO-YBJ; Black crosses: profile of the charged particle density measured by ARGO-YBJ; red squares: charged particle density measured by PRISMA-YBJ; blue triangles: neutron density corrected for the detector efficiency and multiplied by 50 as measured by PRISMA-YBJ. (For interpretation of the references to color in this figure legend, the reader is referred to the web version of this article.)}
 \label{co-example}
 \end{figure}

ARGO-YBJ and PRISMA-YBJ ran  together from January 24 to February 6 , 2013 . After removing two test days, the effective coincident run time was 11 days. During this period  the trigger rate of PRISMA-YBJ was 0.19 $ min^{-1}$. To reduce the sample of coincident event candidates, only showers recorded by ARGO-YBJ  and flagged by  the local trigger of cluster 78 have been considered. This guarantees the detection of showers with high particle density on cluster 78, where the PRISMA-YBJ unit is deployed.  The corresponding trigger rate is 18.2 $min^{-1}$ . The distribution of the time difference $\Delta T = t_{PRISMA}-t_{ARGO}$ is shown in Fig. \ref{selection}, upper plot.  The distribution is not symmetrical and exhibits a large tail  for positive values of $\Delta T$ due to the accuracy of about 10 ms in the  $t_{PRISMA}$ measurement (Sect. 3). The central part of this distribution can be fitted by a Gaussian function with s.d. 0.017 s. A time window of $\pm$ 0.03 s  (about 2 s.d.) around the peak of the distribution has been fixed  to select coincident events  keeping low the number of accidentals. To ensure a good quality of the reconstructed shower parameters, only events with a number of "fired" Big-Pads $>$ 20 at a distance $r$ $<$ 10 m from the shower core position have been selected. 2626 coincident events satisfy  these conditions. The expected number of spurious events is about 110, that is 4.2\% of the total sample.  This low fraction of accidentals makes us confident of  getting a clean correlation between the data recorded by the two arrays. The core position of the selected events is distributed around the PRISMA-YBJ array as shown in Fig. \ref{selection}, lower plot.  On the contrary, the core distribution of the ARGO-YBJ events is almost uniformly spread around the carpet center. This gives an additional evidence of the correlation between the two sets of data. The angular distribution of the coincident events follows closely the angular distribution of the events recorded by ARGO-YBJ (Fig. \ref{angular}) , showing that no particular biases have been introduced in the selection procedure. The number of events with zenith angle less than 30$^{\circ}$ is 1598.  The morphology of a coincident event is shown in Fig. \ref{co-example}. The pattern of a high energy shower detected by the RPC carpet is displayed in the upper plot. The core of the shower hits cluster 78 very close to the PRISMA unit (compare Fig. \ref{co-example}, upper plot with Fig. \ref{array}, lower right plot, where the positions of the EN-detectors are shown) . Particle densities up to about $10^4$ $/m^2$  near the core are recorded by the RPC BigPads. The lateral distribution of charged particles is shown in the lower plot, together with the electron and neutron densities measured by the PRISMA-YBJ detectors at distances less than 4 meters from the impact point of the shower core.

\section{Data analysis and results}

\begin{figure}[t]
 \centering
 \includegraphics[width=0.8\textwidth]{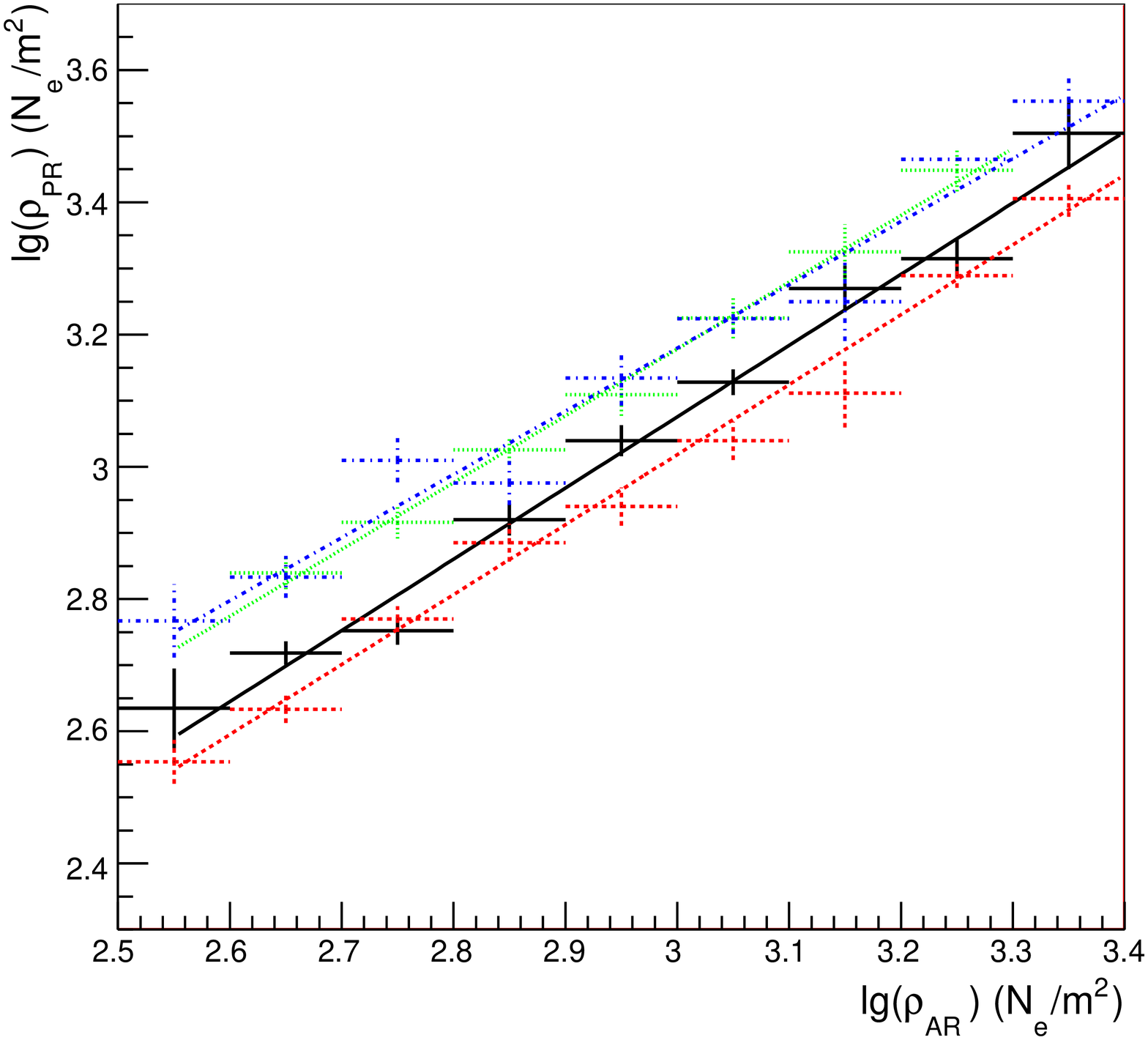}
 \caption{Correlation between the electron densities measured by the four EN-detectors of PRISMA-YBJ via the fast signals ($\rho_{PR}$) and those measured by the corresponding RPCs of ARGO-YBJ ($\rho_{AR}$): D1 (black solid line), D2 (red dashed line), D3 (green dotted line) and D4 (blue dot-dashed line). (For interpretation of the references to color in this figure legend, the reader is referred to the web version of this article.)}
 \label{Nebp-Nepr}
 \end{figure}

The 2626 coincident events have been analyzed to study the performance of the PRISMA-YBJ detectors. Particle densities are measured on the plane orthogonal to the direction of the shower as reconstructed by ARGO-YBJ. Concerning the measurement of the EAS charged component, we have selected the 1598 events recorded at zenith angle less than 30$^{\circ}$. The correlation between $\rho_{PR}$, the electron density measured by each PRISMA-YBJ detector via the fast signals, and $\rho_{AR}$, the electron density measured by each RPC installed beneath, is shown in Fig. \ref{Nebp-Nepr}. To convert the PRISMA-YBJ signals to the corresponding number of minimum ionizing particles the conversion factor obtained in the Moscow laboratory (see Sect. 2) has been adopted. We observe a difference up to 45 \% between the response of the four detectors , well compatible with the laboratory calibration accuracy ($\pm$ 20 \%)  , taking also into account the operation in very different environmental conditions.  The data concerning each detector  have been fitted by means of a straight line, as shown in Fig. \ref{Nebp-Nepr}, obtaining the slopes given in Tab. \ref{table-04}. These slopes are consistent with unity once the ARGO-YBJ RPCs calibration accuracy (5 \% - 10 \%) is taken into account. A weighted average gives 1.05 $\pm$ 0.02. This result indicates a good proportionality of the PRISMA-YBJ  fast signals to the number of charged particles. Thus these data lead us to conclude that a satisfactory performance is achieved for measuring the EAS electron densities once the appropriate absolute calibration is applied.

\begin{table}[h]
\begin{center}
\begin{tabular}{|l|c|}
\hline Detector Number & slope\\ \hline
1 & $1.08 \pm 0.04$ \\ \hline
2 & $1.06 \pm 0.03$ \\ \hline
3 & $1.01 \pm 0.06$ \\ \hline
4 & $0.96 \pm 0.06$ \\ \hline
\end{tabular}
\caption{Slopes of the straight lines of Fig. \ref{Nebp-Nepr}.}
\label{table-04}
\end{center}
\end{table}

Concerning the detection of thermal neutrons, the data obtained with PRISMA-YBJ cannot be directly compared with similar records by ARGO-YBJ since this experiment does not include neutron detectors or other devices sensitive to the EAS hadronic component. Instead, we have reconstructed the neutron lateral density distribution which is strictly correlated to the hadron lateral distribution. These data have been used to find the total number of thermal neutrons that is expected to have a substantial linear correlation with the shower size, since this kind of dependence characterizes the high energy hadrons. Indeed, electrons and hadrons are closely related to each other in the shower development, and a sort of equilibrium turns up \cite{Kempa1}. Many experiments, including the low energy measurements with neutron detectors, prove that the number of hadrons $N_h$ in a shower is almost proportional to the shower size $N_e$, that is
\begin{equation}
N_h =k \times N_e ^{\alpha}
\label{NhNe}
\end{equation}
with $\alpha$ varying between 0.9 and 1.0 \cite{Kempa1}\cite{Grieder}. Thus the total number of thermal neutrons recorded by PRISMA-YBJ can be profitably compared with the shower size measured by ARGO-YBJ, being a quasi-linear correlation one genuine signature of a correct detection of the thermal neutron events.

\begin{figure}[htb!]
 \centering
 \includegraphics[width=0.8\textwidth]{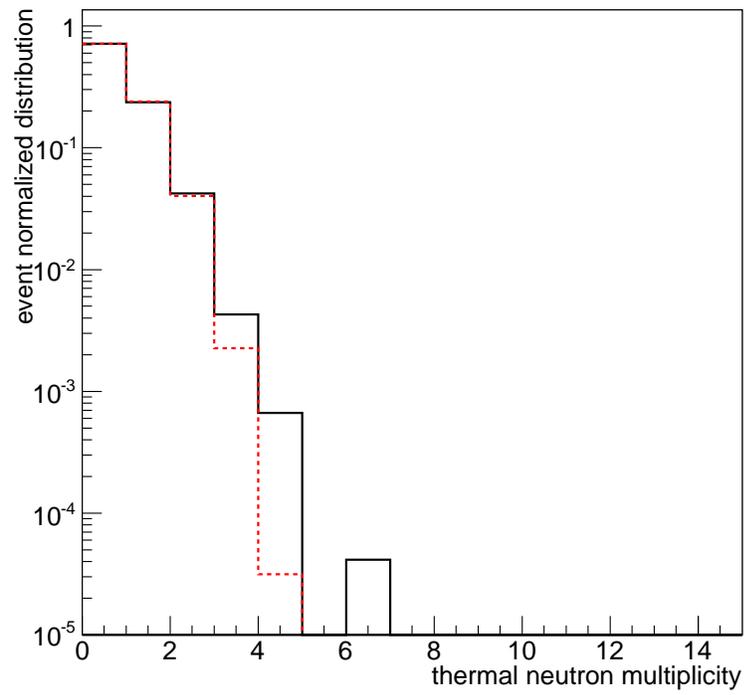}
 \caption{Measured background distribution. Black solid line: data; Red dashed line: Poisson distribution with mean value 0.33. (For interpretation of the references to color in this figure legend, the reader is referred to the web version of this article.)}
 \label{neutronsM0}
 \end{figure}

\begin{figure}[htb!]
 \centering
 \includegraphics[width=0.8\textwidth]{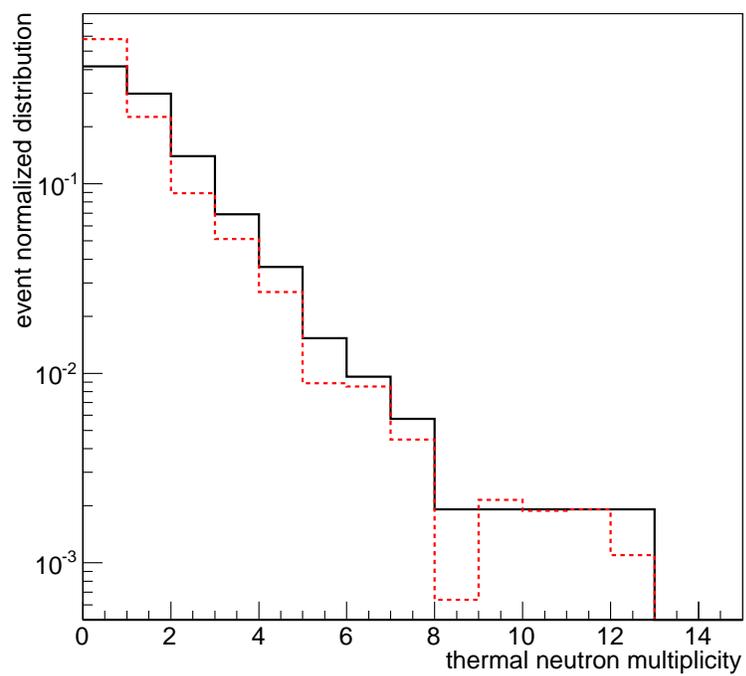}
 \caption{Thermal neutron multiplicity distribution of the matched events with $R$ $<$ 10 m before (black solid line) and after (red dashed line) background correction. (For interpretation of the references to color in this figure legend, the reader is referred to the web version of this article.)}
 \label{neutronsM1}
 \end{figure}

To carry out this analysis, a subset of the 2626 coincident events has been used. Since high energy hadrons are mostly confined around the shower core, only events with reconstructed core location less than 10 m away from the center of PRISMA-YBJ have been selected. Thus, 525 events remain after this selection. 217 events of this sample do not present neutron-like signals in the 10 ms  recording window, while 308 events display neutron-like signals. However, due to the time width of the recording window, some chance signals not associated with air showers could randomly occur. Recognition of chance signals can be performed taking advantage of the statistics recorded by the M0 trigger (see Sect. 3). The probability of one or more chance signals recorded in the 10 ms time window is found to be about 28 \% as shown in Fig. \ref{neutronsM0}, with an average of about 0.33 signals per event. Their multiplicity distribution follows fairly well a Poisson distribution with this mean, as shown in the same figure. Experimental data have been corrected by sampling  spurious signals from this distribution and choosing randomly the involved detectors. After this correction the number of events with neutron-like signals is reduced to 223, with a corresponding total number of
412 detected neutrons. The neutron multiplicity distribution before (black solid line) and after (red dashed line) the background correction is shown in Fig. \ref{neutronsM1}. Finally, selecting events with zenith angle less than 30$^{\circ}$, a sample of 315 coincident events with signal distributions corrected for chance signals is obtained. They are the fraction of the 1598 events with zenith angle less than 30$^{\circ}$ in which the distance between the shower core and the center of the PRISMA-YBJ array is less than 10 m.

\begin{figure}[htb!]
 \centering
 \includegraphics[width=0.49\textwidth]{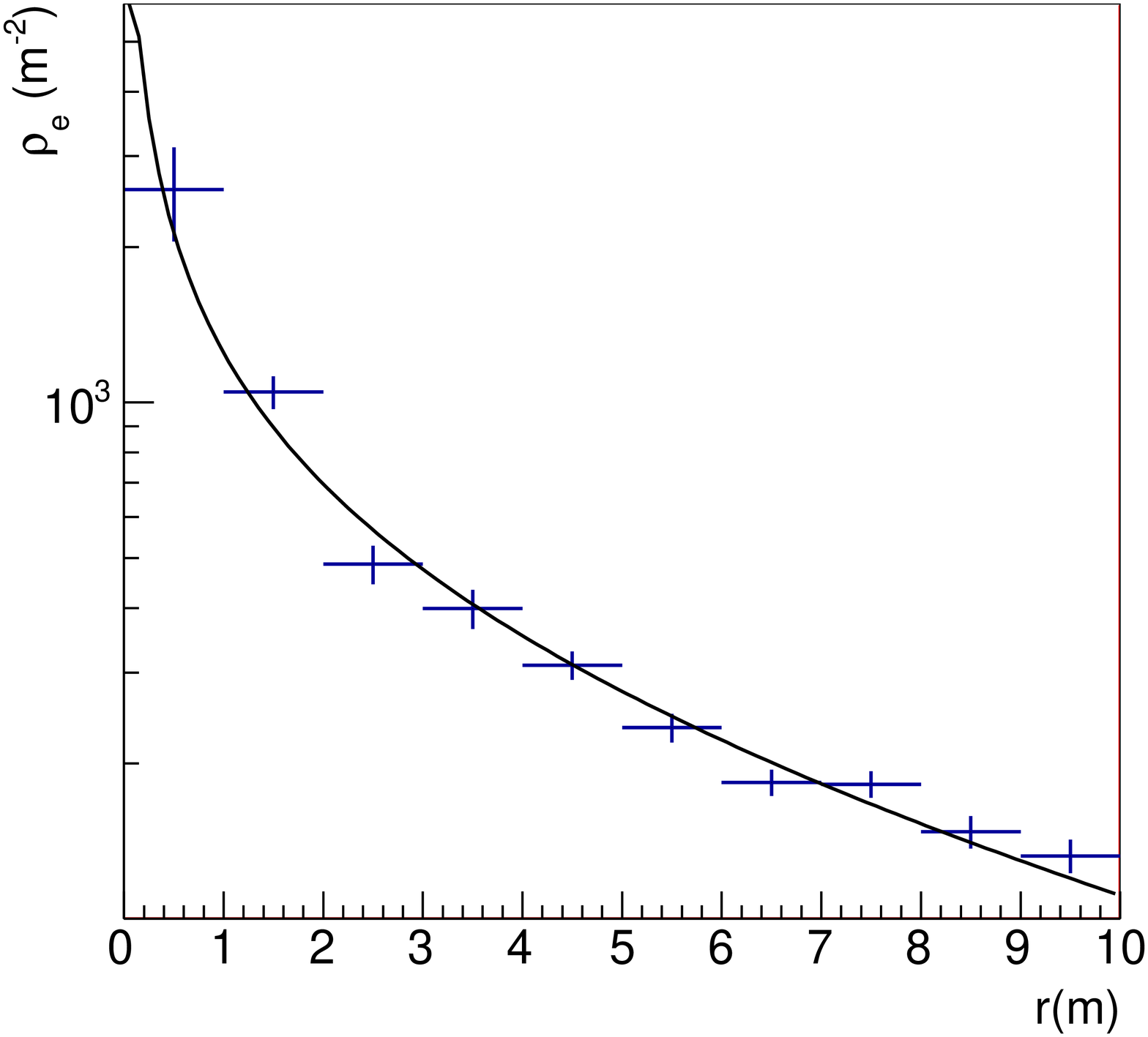}\\
 \includegraphics[width=0.49\textwidth]{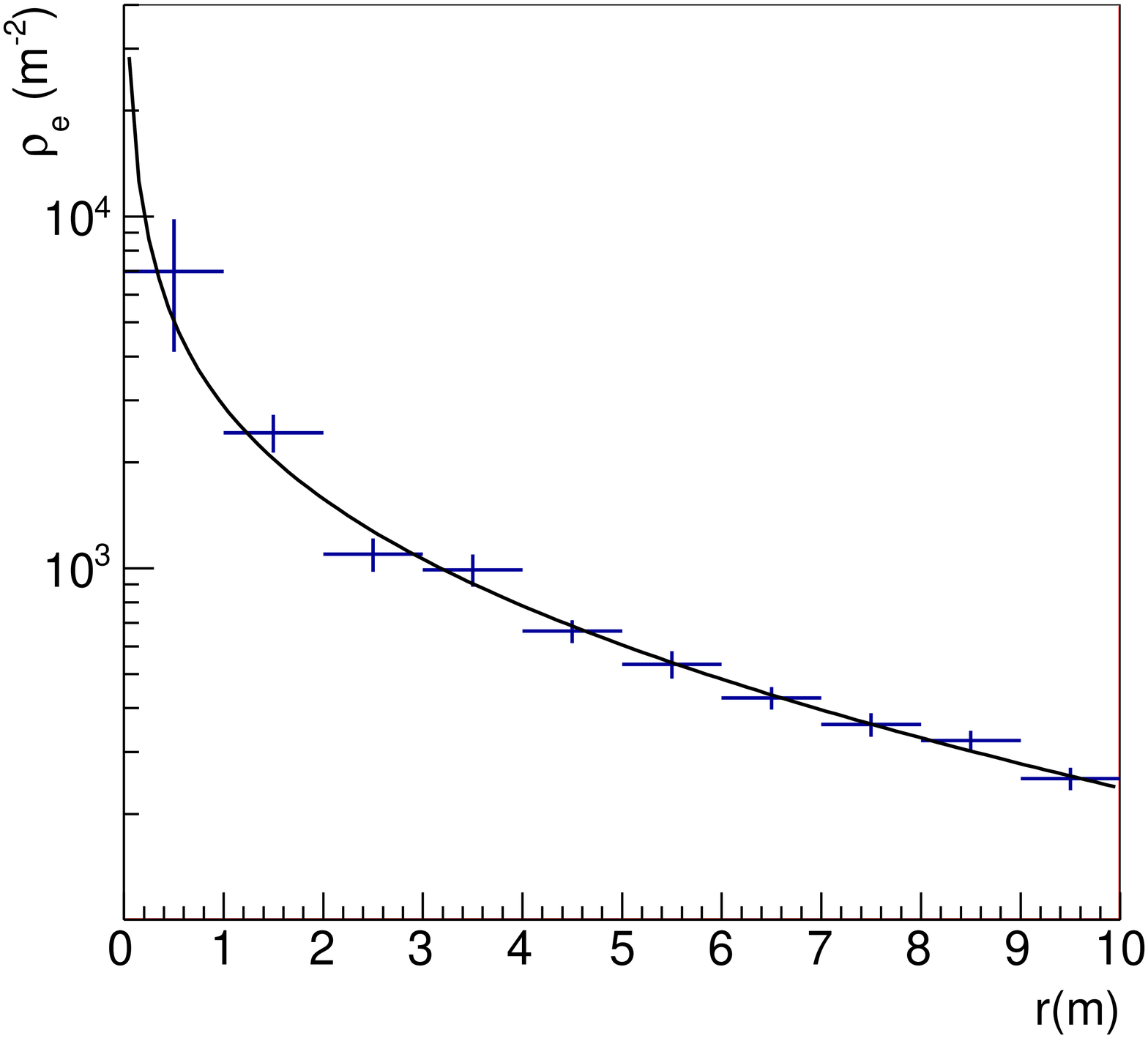}\\
 \includegraphics[width=0.49\textwidth]{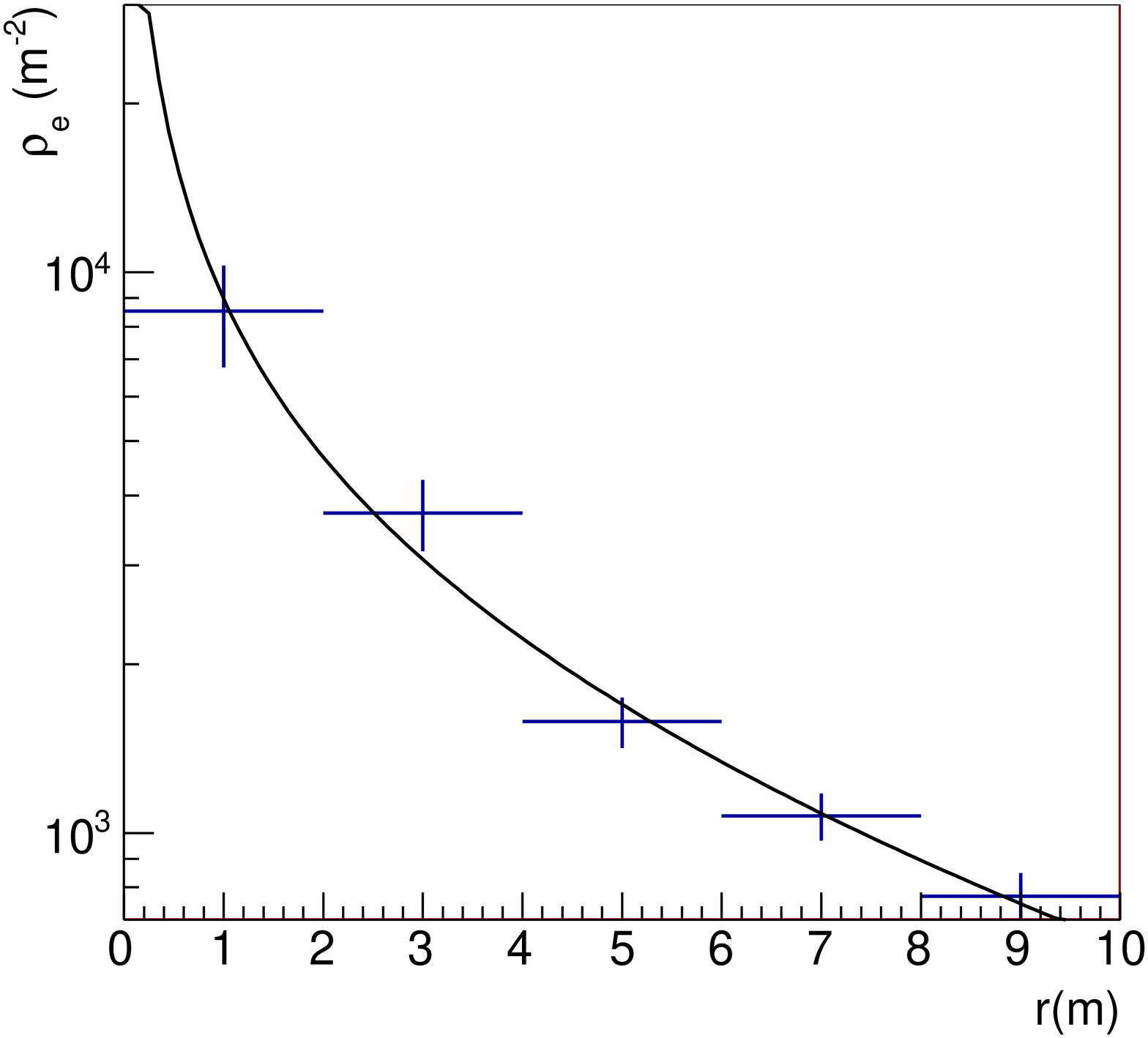}
 \caption{Lateral distributions of electrons measured by PRISMA-YBJ and fitted by the power law function of Eq. (\ref{eLDF}). $r$ (m) is the distance from the shower core, $\rho_{e} (m^2)$ is the electron density. Upper plot: lg($N_{p10}$) $<$ 4.8; Middle plot: 4.8 $<$ lg($N_{p10}$) $<$ 5.4; Lower plot: lg($N_{p10}$) $>$ 5.4}
 \label{elateral}
 \end{figure}

To study the dependence of the neutron yield on the shower size, the events of this sample have been grouped according to the truncated size $N_{p10}$, that is the number of particles measured by ARGO-YBJ within a distance of 10 m from the shower axis, a quantity  not biased by effects due to the finite detector size. Three different size intervals have been chosen selecting 185 events with lg($N_{p10}$) $<$ 4.8 , 107 events with 4.8 $<$ lg($N_{p10}$) $<$ 5.4 and 23 events with lg($N_{p10}$) $>$ 5.4 . The relation of the truncated size with the energy for each primary cosmic ray component  is given in \cite{Iacovacci}. Typically, the first size group is associated to primary protons (irons) of energy below 550 TeV (1.2 PeV), the second group to primaries of energy around 1 PeV, while the showers belonging to the third group are mainly initiated by cosmic rays with energies greater than 1.5 PeV. The electron density distribution of these events as measured by PRISMA-YBJ is shown in Fig. \ref{elateral}. Data can be fitted very well  by means of the functional form (a NKG-like formula) currently used to describe the ARGO-YBJ measurements \cite{Bernardini}:
\begin{equation}
\rho_e (r) = A \times (\frac{r}{r_m})^{s'-2} \times (1+\frac{r}{r_m})^{s'-4.5}
\label{eLDF}
\end{equation}
where $r$ (m) is the distance from shower cores, $\rho_{e}$ (m$^2$) is density of electrons, $A$ is a normalization factor, $s'$ is the shape parameter that plays the role of the lateral age, and $r_m$ is a constant scale radius which is 30 m. The fit parameters are given in Tab. \ref{table-eLDF}. It shows again the full consistency among the  electron measurements performed by the two detectors. It is interesting to note that the lateral age parameter $s'$ appearing in this formula runs from 1.31 for the first size group to 1.20 for the third size group. It is a shape parameter strictly related to the shower age $s$ which measures the stage of the longitudinal shower development. According to the relations reported in \cite{Bernardini}, the age parameter $s$ results about 1.14 for the higher energy showers used in the present analysis, indicating that these showers are sampled not far from their maximum, about 100 g/cm$^2$ above the Yangbajing atmospheric depth.

\begin{table}[h]
\begin{center}
\begin{tabular}{|l|c|c|c|}
\hline $N_{p10}$ intervals & $\chi^2/ndf$ & $A$ (m$^{-2})$ & $s'$ \\ \hline
lg($N_{p10}$) $<$ 4.8         & 15.9/8 & $130 \pm 9 $ & $1.30 \pm 0.04$ \\ \hline
4.8 $<$ lg($N_{p10}$) $<$ 5.4 & 6.34/8 & $270 \pm 22$ & $1.27 \pm 0.05$ \\ \hline
lg($N_{p10}$) $>$ 5.4         & 2.09/3 & $687 \pm 93$ & $1.21 \pm 0.08$ \\ \hline
\end{tabular}
\caption{Fit results of the distributions shown in Fig. \ref{elateral}.}
\label{table-eLDF}
\end{center}
\end{table}

\begin{figure}[htb!]
 \centering
 \includegraphics[width=0.49\textwidth]{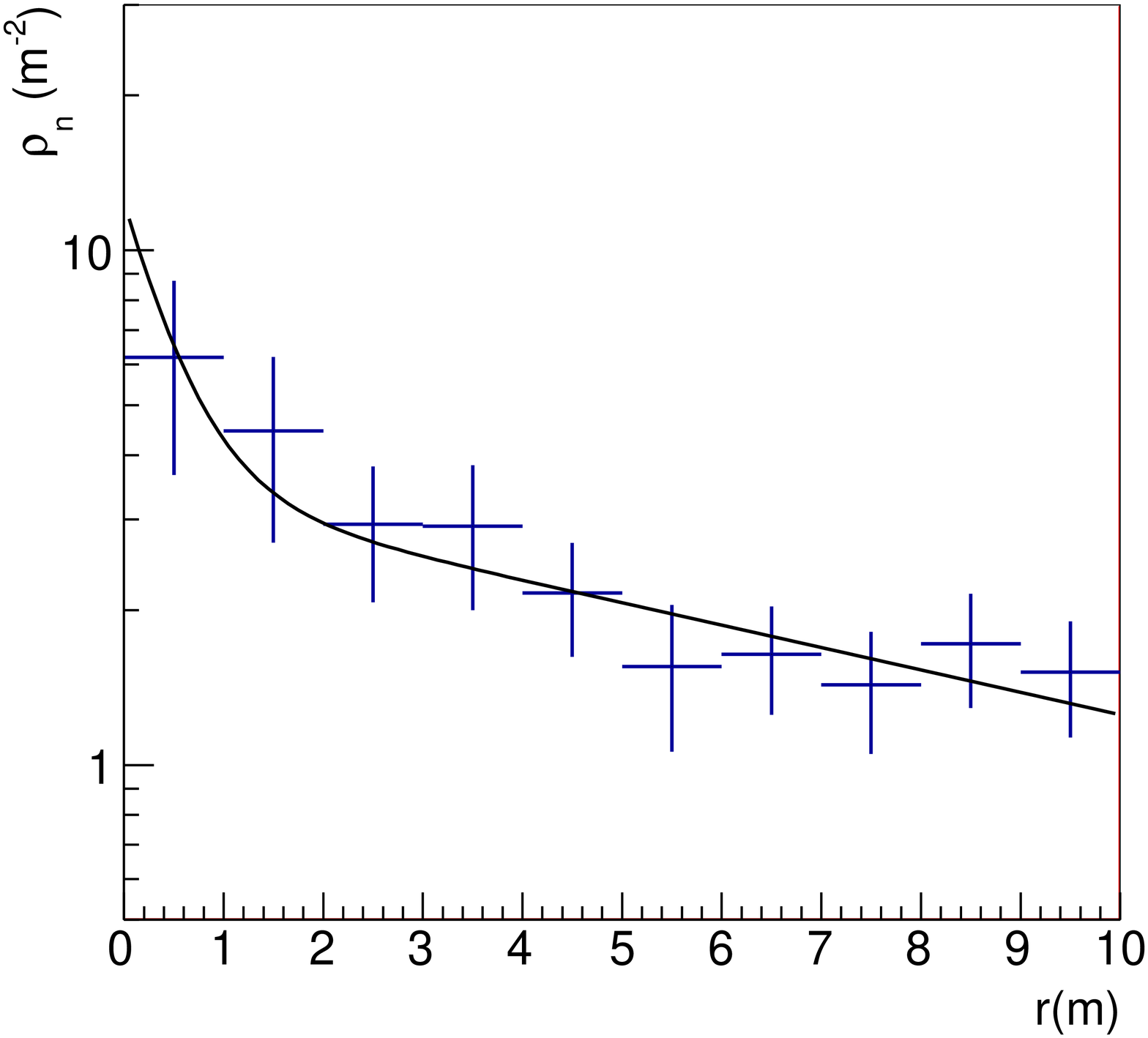}\\
 \includegraphics[width=0.49\textwidth]{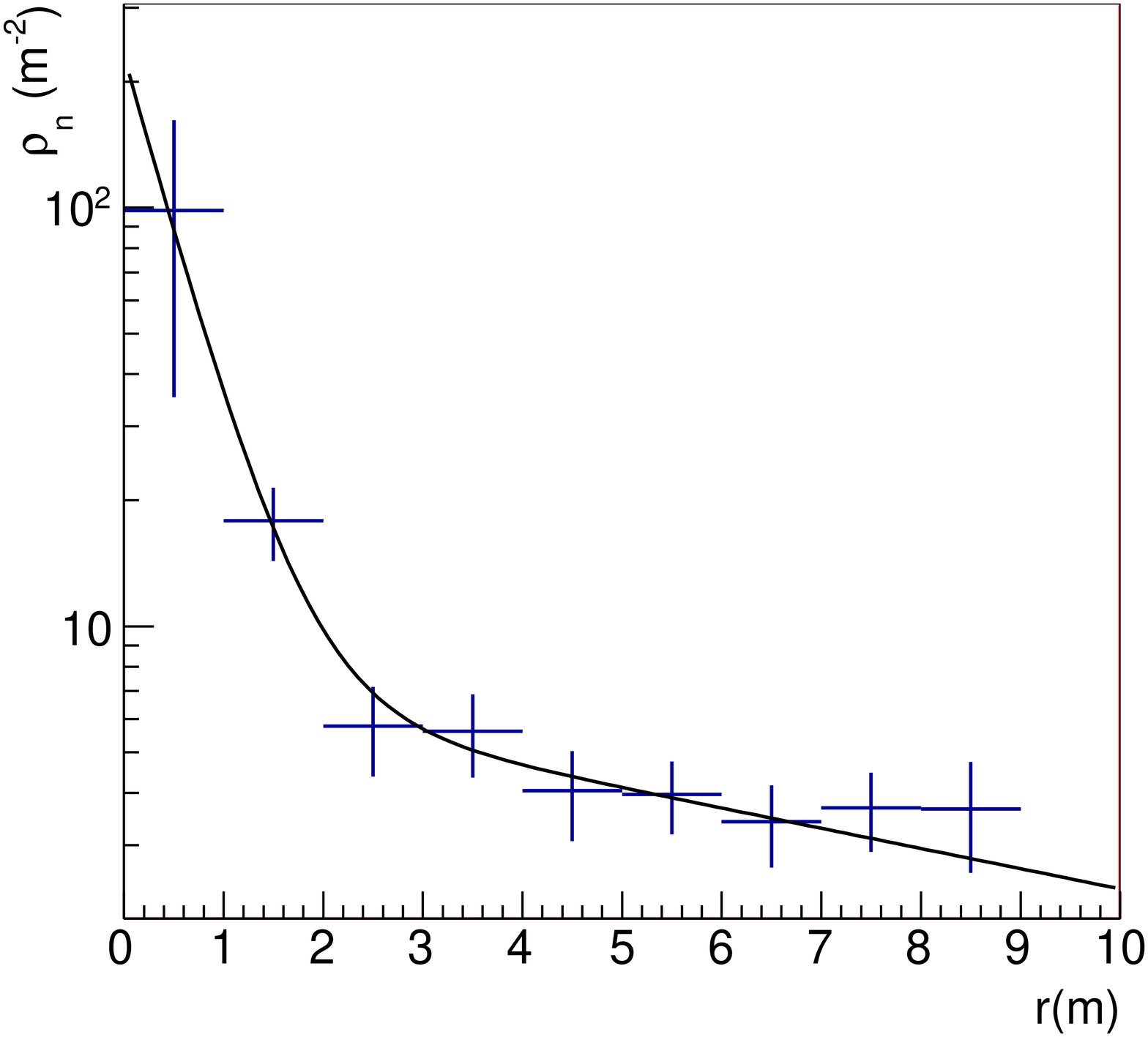}\\
 \includegraphics[width=0.49\textwidth]{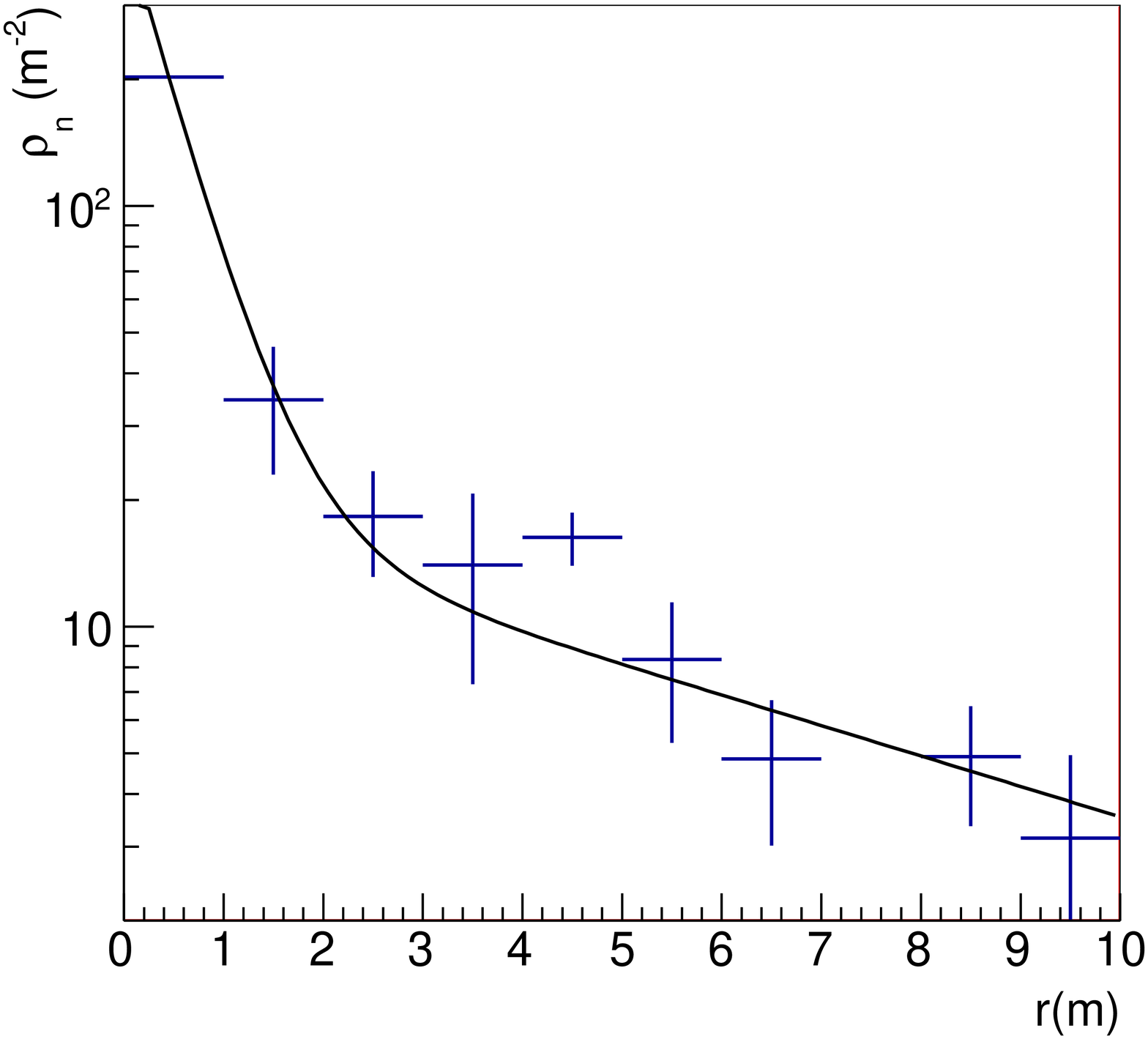}
  \caption{Lateral distributions of thermal neutrons fitted by the double exponential function of Eq. (\ref{nLDF}). $r$ (m) is the distance from the shower core, $\rho_{n} (m^2)$ is the neutron density. Upper plot: lg($N_{p10}$) $<$ 4.8; Middle plot: 4.8 $<$ lg($N_{p10}$) $<$ 5.4; Lower plot: lg($N_{p10}$) $>$ 5.4.}
 \label{lateral}
 \end{figure}

The neutron lateral distributions, that is the average thermal neutron density per event as a function of the distance $r$ from the shower core, are reported in Fig. \ref{lateral} for each group. To obtain this result, we have measured for each event the distance of the EN-detectors from the shower core and we have incremented the corresponding bin of the $r$-profile with the content of each EN-detector. Data have been normalized to the total number of events recorded in each size group, including events without neutron signals, and corrected for the detector efficiency. The shape of these distributions is easily related to the involved physics processes. Experimental data taken at high altitudes  show that the hadron distribution at EAS core follows an exponential behavior \cite{Sreekantan}. The hadron interaction and thermal neutron production have been studied using a GEANT4 based simulation package \cite{PRISMA12}. The results of the simulation show that the lateral distribution of these thermal neutrons can be conveniently described by a two exponential functional form. Indeed we found that a function of this type,
\begin{equation}
\rho_n (r) = \rho_{0} \times e^{-(r/r_{0})} + \rho_{1} \times e^{-(r/r_{1})}
\label{nLDF}
\end{equation}
provides a reasonable fit to our data (Fig. \ref{lateral}), with $r_{0}$ = 0.5 $\pm$ 0.1 m  and $r_{1}$ slowly decreasing with increasing shower size , that is  $r_{1}$ = 10.0 $\pm$ 1.0 m , 9.0 $\pm$ 1.0 m, 6.0 $\pm$ 1.0 m for the three size groups. The fit parameters are given in Tab. \ref{table-nLDF}. A simple explanation of this behavior assumes that neutrons close to the shower axis are produced by the interaction of hadrons with the soil around the detector, while more sparse neutrons come from the shower development in air and from the interaction with the building roof.

\begin{table}[h]
\begin{center}
\begin{tabular}{|l|c|c|c|}
\hline $N_{p10}$ intervals & $\chi^2/ndf$ & $\rho_0 (m^{-2})$ & $\rho_1 (m^{-2})$ \\ \hline
lg($N_{p10}$) $<$ 4.8         & 2.44/8 & $9.0 \pm 6.8$ & $3.41 \pm 0.32$ \\ \hline
4.8 $<$ lg($N_{p10}$) $<$ 5.4 & 2.69/7 & $222 \pm 65 $ & $7.17 \pm 0.65$ \\ \hline
lg($N_{p10}$) $>$ 5.4         & 20.1/7 & $456 \pm 230$ & $18.7 \pm 2.3 $ \\ \hline
\end{tabular}
\caption{ Fit results of the distributions shown in Fig. \ref{lateral}.}
\label{table-nLDF}
\end{center}
\end{table}

\begin{figure}[htb!]
 \centering
 \includegraphics[width=0.8\textwidth]{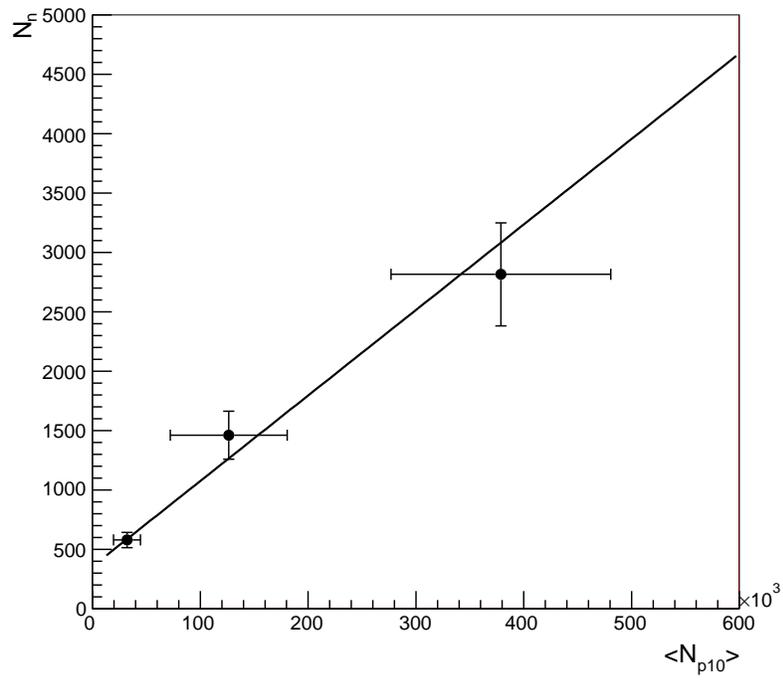}
 \caption{Total number of thermal neutrons $N_n$ measured by PRISMA-YBJ within $r<$ 10 m vs. the mean value of $N_{p10}$ measured by ARGO-YBJ. A linear function (intercept = $356\pm 144$, slope = $(7.2\pm 2.2)\cdot 10^{-3}$) fits properly the experimental data.}
 \label{Nn_Np10}
 \end{figure}

Integrating the neutron lateral distribution up to $r$ = 10 m we obtain the total number of thermal neutrons around the shower axis. As shown in Fig. \ref{Nn_Np10} , the total number of thermal neutrons measured by PRISMA-YBJ is in a good linear relation with the truncated size measured by ARGO-YBJ, confirming the correct operation of the PRISMA-YBJ detector to measure thermal neutrons associated to air showers. A detailed description of this relation requires the use of an accurate simulation that is beyond the scope of the present paper. However, it is worth to note the rich thermal neutron yield, as expected from the fact that their production is initiated also by low energy hadrons (see Eq. \ref{nE} in Sect. 1). From Fig. \ref{Nn_Np10} we find that the ratio of the thermal neutron number to the electron number within 10 m from the shower axis is about $7 \times 10^{-3}$ , at least one order of magnitude greater than the ratio of the hadron number to the electron number \cite{Kempa2}. Higher statistics and comparison with predictions of dedicated simulations will be clearly necessary to better establish the exact nature of the physical correlations detectable by the measurement of thermal neutrons in air showers. However, the present result, as it is, shows that, despite of its limited efficiency, the EN-detector is a very attractive device to estimate the hadron content in showers originated by cosmic rays of energy greater than 1 PeV.

\section{Conclusions}

This paper is addressed to the use of the EN-detector in the framework of EAS measurements . The sensitive layer of this detector is basically made of an alloy of an inorganic ZnS(Ag) scintillator alloyed with LiF enriched with the $^{6}$Li isotope. Its originality lies in simultaneously recording the air shower charged component and the thermal neutrons generated by high energy hadrons. This is achieved by exploiting the shape and timing differences of the signals induced by these particles. The detection of thermal neutrons by means of a quite simple device open a new opportunity to estimate the hadron content in EAS. Their observation has been so far plagued by the need of building huge calorimeters. Thermal neutrons in EAS generated by high energy primaries above 100 TeV have been detected at high altitude. This result has been accomplished by operating a PRISMA unit of four EN-detectors in coincidence with the air shower experiment ARGO-YBJ located at the Cosmic Ray Observatory of Yangbajing (Tibet, 4300 m a.s.l.). With the help of the reconstructed shower parameters from ARGO-YBJ a calibrated set of data has been obtained. The response of the EN-detectors to the EAS charged particles exhibits a good linearity. The thermal neutrons are found distributed around the shower core with a very narrow lateral shape and their total number is well correlated with the truncated shower size measured by ARGO-YBJ. Both features are consistent with the ones characterizing the EAS high energy hadrons. The present analysis concentrates only on experimental findings, while a comparison with detailed Monte Carlo simulations is deferred to a future publication. However, the analysis of a sample of more than two thousand EAS events confirms that the EN-detectors worked properly at high altitude in combination with an array of particle detectors.

With the growing interest for high energy cosmic ray research involving very large surface detectors, the possibility of measuring thermal neutrons and the charged component with an unique detector deployed over a large area makes this device an attractive and promising cost effective tool for future apparatuses. Indeed, it has been proposed to deploy the PRISMA units in the planned LHAASO array \cite{LHAASO1}. The operation of the EN-detectors in the LHAASO experiment will enable the estimate of the hadron content in EAS, increasing its capability to determine the energy and nature of high energy cosmic rays.

\section{Acknowledgment}

This work is supported in China by NSFC (No. 10120130794, No.10975046, No. 11205165, No. 11375210, No.11375052), the Chinese Ministry of Science and Technology, the Chinese Academy of Sciences, the Key Laboratory of Particle Astrophysics, CAS, and in Italy by the Istituto Nazionale di Fisica Nucleare (INFN). We also acknowledge the essential support of W. Y. Chen, G. Yang, X. F. Yuan, C. Y. Zhao, R. Assiro, B. Biondo, S. Bricola, F. Budano, A. Corvaglia, B. D¡¯Aquino, R. Esposito, A. Innocente, A. Mangano, E. Pastori, C. Pinto, E. Reali, F. Taurino, and A. Zerbini in the installation, debugging, and maintenance of the detector. Authors acknowledge financial support from RFBR (grants 14-02-00996 and 13-02-00574), RAS Presidium Program ¡®Fundamental Properties of Matter and Astrophysics¡¯.









\newpage

\end{document}